\documentclass[]{interact}

\usepackage{hyperref}
\hypersetup{
    colorlinks=true,
    citecolor={black},
     linkcolor=black,
filecolor=black,      
    urlcolor=blue,
  pdftitle={Overleaf Example},
  pdfpagemode=FullScreen,
    }
    
\newcommand{\rd}{\textcolor[rgb]{1.00,0.00,0.00}}    
    
\usepackage{amssymb}
 \usepackage{array}
\usepackage{comment}
\usepackage{xcolor}
\usepackage{epsfig}
 
\usepackage{algpseudocode}
\usepackage{rotating}
\usepackage{subfigure}
\usepackage{amsmath}
 \usepackage[linesnumbered,ruled]{algorithm2e}
  \SetKwRepeat{Do}{do}{while}%
\usepackage{float}

\usepackage{epstopdf}
\usepackage{subfigure}

\usepackage{natbib}
\bibpunct[, ]{(}{)}{;}{a}{}{,}

\theoremstyle{plain}

\theoremstyle{definition}

\theoremstyle{remark}

\usepackage{setspace}
\begin{document}


\title{Dynamic Risk-Adjusted Monitoring of Time Between Events: Applications of NHPP in Pipeline Accident Surveillance}

\author{
\name{Hussam Ahmad\textsuperscript{a}, Adel Ahmadi Nadi\textsuperscript{b},  Mohammad Amini\textsuperscript{a} and Subhabrata Chakraborti\textsuperscript{c}\thanks{CONTACT Adel Ahmadi Nadi. Email: adel.nadi67@gmail.com}}
\affil{\textsuperscript{a}Department of Statistics, Faculty of Mathematical Sciences,  Ferdowsi University of Mashhad,  Mashhad, Iran; \\\textsuperscript{b}Department of Statistics and Actuarial Science, University of Waterloo, Waterloo, Canada;\\ \textsuperscript{c} Department of Information Systems, Statistics and Management Science, The University of Alabama, Tuscaloosa, Alabama, USA.}
}

\maketitle

\begin{abstract}
Monitoring time between events (TBE) finds widespread applications in industrial settings. Traditional Statistical Process Monitoring (SPM) methods often assume TBE variables follow an exponential distribution, implying a constant intensity function for the failure process. While this assumption may be suitable for products with homogeneous quality, it becomes less tenable for complex products, such as repairable systems, where failure mechanisms evolve with time due to degradation or aging. In such cases, the Non-Homogeneous Poisson Process (NHPP), which allows for time-varying failure intensity, is a more suitable model.  Moreover, the failure patterns in such complex systems are often influenced by risk factors, such as environmental conditions or human interventions. Additionally, system failures of this nature entail costs for restoration.  This work proposes a novel approach: a risk-adjusted control chart based on the NHPP model, specifically designed for monitoring the ratio, Cost/TBE, called the average cost per time unit (AC). The efficacy of the proposed method is evaluated through a series of simulations. The numerical study attests to the favourable performance of the developed chart. The numerical study shows the outperformance of the developed chart.  The proposed chart is applied to monitor pipeline accidents over time, considering the influence of various risk factors. 
\end{abstract}


\section{Introduction}
 
An important focus of Statistical Process Monitoring methods is to develop efficient control charts to monitor the quality of a process over time, and to improve performance by providing valuable insights into process stability and variability (\cite{Montgomery}). Though the ideas were first introduced to monitor industrial processes (e.g., a manufacturing process), they have now gained a strong foothold in many nonindustrial sectors, including medicine, healthcare and public health, surveillance, financial markets, climate and environment, chemical analysis, network monitoring, and change-point problems, see for example \citep{You2020, Stevens2021, Bisiotis2022,  Qiu2022}. In an industrial context, SPM techniques can noticeably contribute to improving the quality of manufacturing processes by monitoring the output of the process over time. As another example in a financial context, SPM tools can be used for various financial applications such as portfolio monitoring and stock trading (\cite{Kovarik2015}). \\
 
One important application of SPM is in the context of monitoring defectives or failures, where attribute control charts (e.g. $p$ chart) have been traditionally used. However, for monitoring high-quality processes, where the number of defectives (failures) is often most likely to be zero, a more efficient monitoring procedure is to focus on the elapsed time between two successive defective items (failures), and not on monitoring the number of defectives. Time-between-events (TBE) control charts are widely used for these processes. Note that in this context, although traditionally, the event of interest has been ``producing a defective item", the idea is not limited to this case, and the event of interest can include a broad range of applications, even in non-manufacturing contexts, such as the time between disease outbreaks in a health care context, or the time between failures of operating systems in the industrial sector. To be consistent throughout the paper, ``system" may refer to a ``manufacture item",  a ``production process", or a ``repairable system". Likewise, a ``failure or defective"  will be used to refer to any event of interest.  \\

Most of the studies on TBE charts assume that failures occur according to a homogeneous Poisson process (HPP), which results in the distribution of TBE being the well-known exponential distribution \citep{Castagliola, Kumar2022, Ahmad1, Ahmad2}.  This means that the deterioration mechanism of the production equipment, say its hazard rate function, is assumed to be constant over time. While this assumption may apply in some cases, it may not be valid even for non-complex production equipment where deterioration is expected with time. This argument could also be true in other monitoring settings. For example, if the goal is to monitor the real-time condition of a repairable system via monitoring the time between its failures, then the HPP may not be an efficient choice. A repairable system is a system that can be repaired after breaking down, thus its failure rate is most likely to change with time and accordingly the non-homogeneous Poisson process (NHPP) is a more realistic choice to model the number of failures of such a system \citep{Guler}. In practice, many systems exhibit complexities and dynamic behaviours, making it challenging to assume a constant failure rate. In such cases, the NHPP offers a more accurate representation of such systems by accounting for varying rates and capturing different failure patterns \citep{Pradhan, Hong}.  That is why the NHPP plays a central role in modelling the failure counts of complex systems in reliability analysis and maintenance literature \citep{Xu2017, Cha2018, Safaei2019, Safaei2022}. Despite this, the attention to NHPP has been comparatively low in the SPM literature. \\

It is essential to recognize that any failure of a system may come with costs to bring it back to operation. These costs may include expenses for repairs and replacements, parts and labour, downtime, warranty claims, as well as potential penalties and fines, either individually or in combination. Since to total cost (TC) associated with each failure may be affected by various factors, it makes more sense to treat it as a random quantity. In the TBE literature,  the ``amplitude" variable is used as a representative of the random TC. Many techniques have been developed for monitoring the TBE and amplitude/TC variables simultaneously that are also known as the TBEA control charts \citep{2.1, 2.2, 2.3}. \\

The failure time datasets often include information about some risk factors (also known as covariates or confounding variables) in addition to the time to failures and costs. These factors can noticeably affect the performance of the system. They encompass a wide range of information and ignoring them in the analysis can cause a high degree of variation and overall deterioration in the system's performance \citep{Zheng, Zhang}. There are several models in the literature to incorporate the relevant risk factor information. These include the well-known Cox proportional hazards model, which is more likely to be applied to account for system-reliability-related risk factors in an industrial context, and the accelerated failure time model which is more commonly used to account for health-related risk factors in a clinical context (\cite{Steiner2010}). Applying a proper risk model to account for the effect of risk factors provides a more informative framework \citep{Cockeran}. In the healthcare-related SPM domain, several control charts have been developed that account for the effect of patient's preoperative risk factors. Among the pioneering works in this area, \citet{Steiner2000, Steiner2001} studied risk-adjusted control charts to track patient death rates. Since then, a huge body of risk-adjusted monitoring methods has been proposed by different scientists \citep{Steiner2014,Woodall2015, Sachlas2019}. Also, \citet{Paynabar} introduced a comprehensive phase I risk-adjusted control chart for monitoring binary surgical outcomes by considering categorical covariates. \cite{Steiner2014a} provides a list of justifications for risk adjustment in the health context. For example, they stated that patients may have different conditions before treatment and thus are not expected to be homogeneous (unlike manufactured parts). One can argue the same is true in the case of monitoring the performance of systems in the presence of different risk factors such as material, labour, weather conditions, and so forth. Risk adjustment in this case takes care to distinguish the failure mechanism of the system itself from the risk model. By doing this, the monitoring method would be more sensitive to the changes of system failures and associated costs \citep{Tian, Mun}.\\

The main objective of this paper is to develop a risk-adjusted control chart for monitoring the average cost per time unit, defined as the ratio TC/TBE, using the NHPP as the underlying stochastic process for the failure counts. The adoption of the NHPP model offers a crucial advantage, as it allows the failure intensity of the system to change with time. Additionally, the incorporation of the total cost and risk factors directly into the control chart analysis represents a novel approach to monitoring procedures. The rest of the paper is structured as follows: Section \ref{sec:ME} presents a motivational example based on pipeline accident data that forms the background of this work. Section \ref{NHPP} briefly introduces the NHPP with two widely used intensity models, as well as discusses the risk adjustment approach. Section \ref{sec:copula} offers an overview of the basic principles and background of copula theory. Moving on, Section \ref{sec:control.chart} focuses on the construction of the proposed control chart, while Section \ref{sec:Per.Ev} presents a numerical study that evaluates the performance of the charts. In Section \ref{sec:Example}, a practical application of the proposed charts based on real data examples is discussed. Finally, the study is concluded in Section \ref{sec:Con} with a concise summary of the findings, and conclusions drawn.

 \section{Motivational Example}\label{sec:ME}

The Pipeline and Hazardous Materials Safety Administration (PHMSA) of the United States Department of Transportation has released a comprehensive report, including a dataset on oil pipeline accidents between January 1, 2010, and September 1, 2017.  The accident/failure of oil pipelines refers to an incident in which the pipeline leaks or ruptures, releasing the contents of the pipeline. Such failures can be hazardous and can potentially cause damage to the environment and human health as well as economic losses \citep{LU, Xi2023, Wang, Li2016}.  The financial consequences of pipeline accidents are also substantial \citep{White}. The total cost of oil spill response and cleanup operations globally ranges from tens of millions to billions of dollars annually \citep{etkin}. That is why it is crucial to provide efficient strategies to inspect the pipelines, monitor their potential failures, and develop emergency response plans in case of a failure. All these kinds of efforts could be highly beneficial to minimize the risk and cost of failures, as well as to prevent any unexpected failures. By closely monitoring pipeline accidents, operators can detect potential issues at an early stage and take corrective and preventive actions before they either escalate or happen again in the future.

 The database comprises details of 2,789 accidents and is publicly available on the Kaggle website (https://www.kaggle.com/datasets/usdot/pipeline-accidents).  Various costs (e.g., property damage costs, lost commodity costs, emergency damage costs, etc.) were incurred in order to repair the pipeline and bring it back into the network. In addition, the dataset provides the TC associated with each accident, which is the aggregate sum of all incurred costs. Furthermore, at the time of each accident, information regarding the pipeline type and location, the type of hazardous liquid involved, the accident location, and the accident cause has also been recorded. All these risk factors could potentially affect the intensity or pattern of the accidents. 
 
A number of articles on the analysis of pipeline failures show that an NHPP provides a satisfactory model for pipeline incidents, for example \citep{cobango2016, Kermanshachi2020}. Accordingly, an NHPP that accounts for a time-dependent accident rate and adjusts for risk factors is expected to provide a better statistical model for the accident mechanism of pipelines. In this context, there is a need to develop a monitoring technique based on a risk-adjusted NHPP model. Thus, the proposed monitoring approach can serve as a real-time surveillance system to detect statistically and practically significant (unusual) deviations (increases and decreases) in the times between successive accidents and the associated total cost from the baseline of (expected) accident patterns. Accordingly, the proposed methodology can help in dealing with and mitigating these undesirable, and unsafe situations and manage the associated risks and costs.

\begin{figure}[H]
 \centering
 \includegraphics[scale=.9]{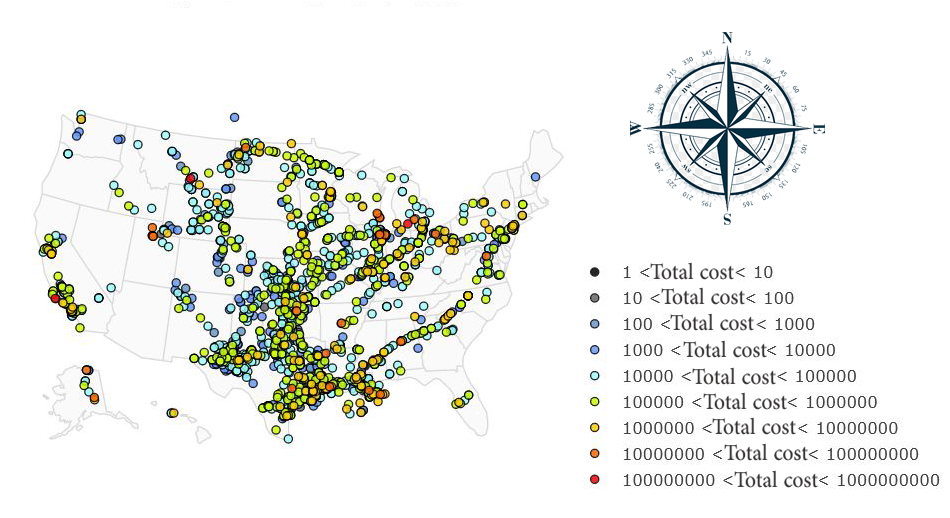} 
 \caption{The distribution of the accidents across the US along with TC in USD.}
\label{fig:map} 
 \end{figure}

\begin{figure}[H]
 \centering
 \includegraphics[scale=0.65]{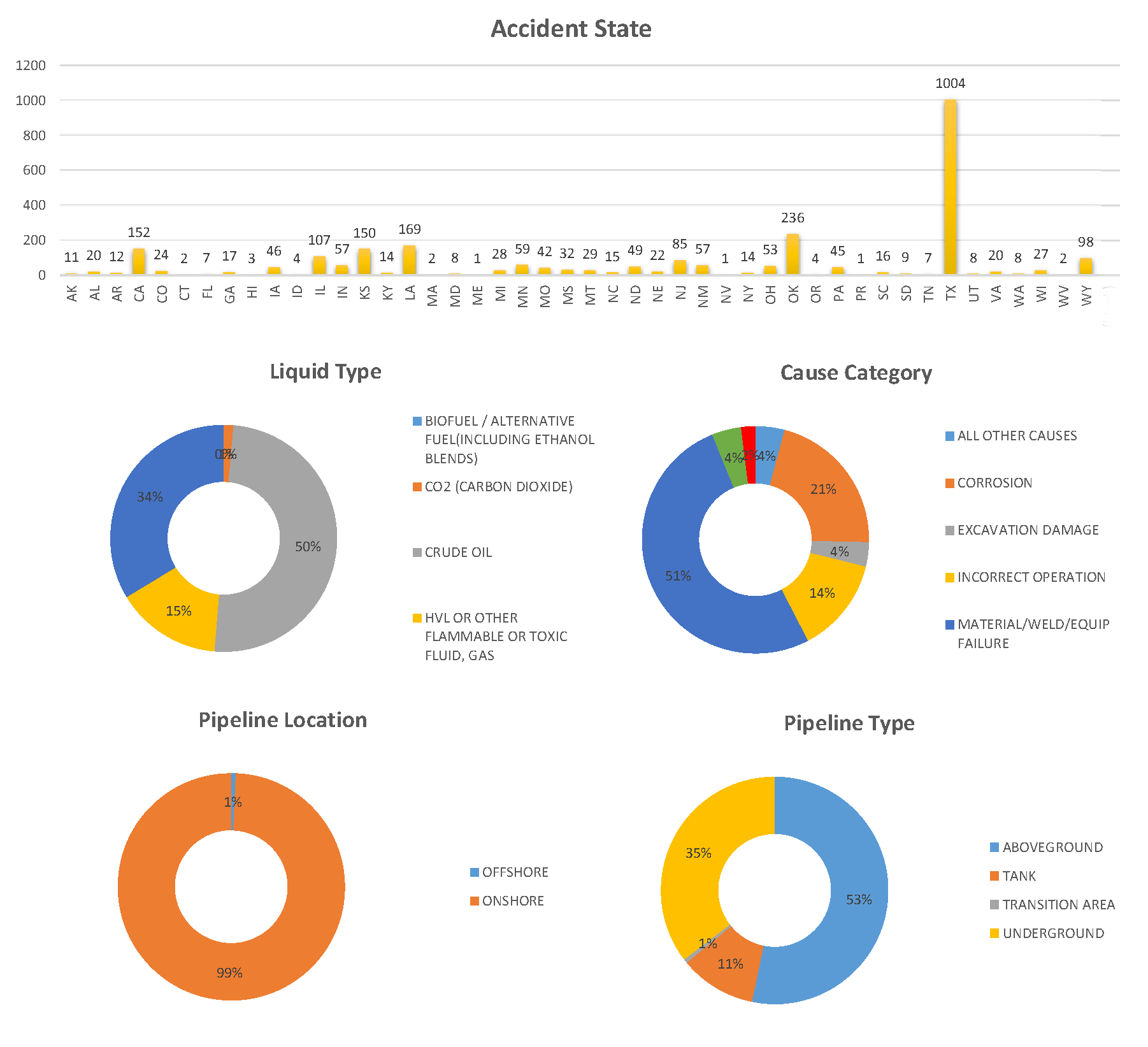} 
\caption{The distribution of accidents across five risk factors.} \label{fig:Dist-Data}
\end{figure}

In what follows, we present additional information about the dataset to provide a deeper understanding. Note that the majority of these incidents are concentrated in the central and southern regions of the United States (US). This geographical distribution of accidents across the US is depicted in Figure \ref{fig:map}. For instance, the state of Texas has reported 1004 accidents, whereas the number of accidents is considerably lower in the northwest of the US, such as in the state of Oregon state, where only four accidents have been recorded (see Figure \ref{fig:Dist-Data}). According to Figure \ref{fig:Dist-Data}, approximately 50$\%$ of accidents occur in pipelines that are used for transporting crude oil. However, when considering the location of the pipelines, 99$\%$ of the accidents happen with onshore pipelines, with 53$\%$ of them being above ground.  Additionally, about 51$\%$ of accidents are attributed to material, welding, or equipment failures.

\begin{figure}[H]
\centering
\subfigure[TBE (in hour).]{%
\resizebox*{7cm}{!}{\includegraphics[width=5cm,height=5cm]{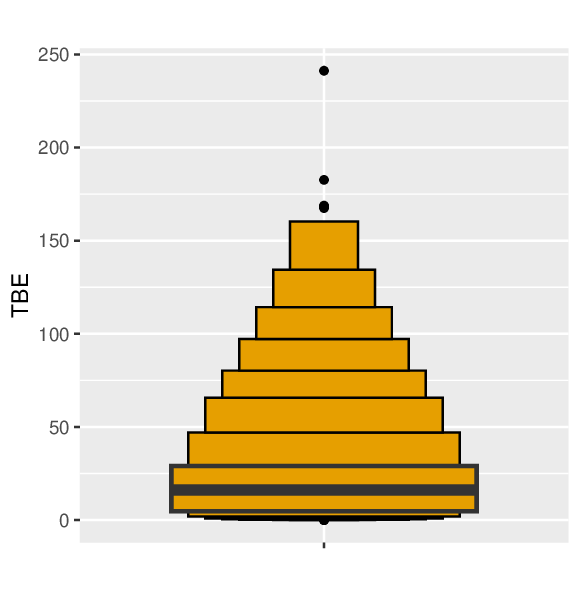}}}
\subfigure[TC (in million USD).]{%
\resizebox*{7cm}{!}{\includegraphics[width=5cm,height=5cm]{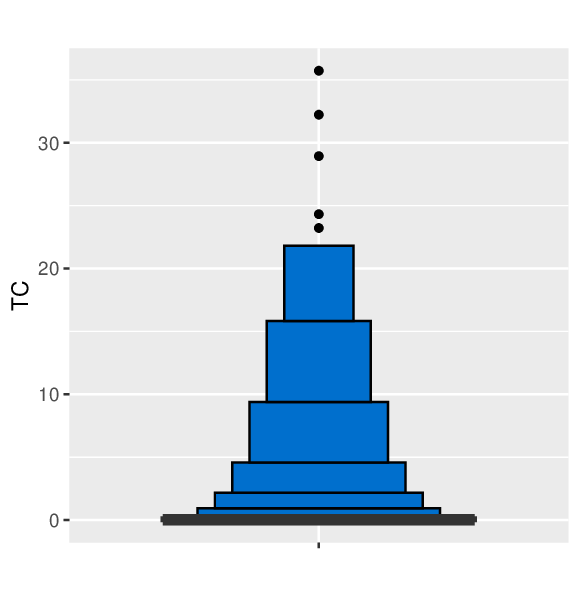}}}
\caption{Letter-value plots for TBE and TC.} \label{fig:TBE-cost}
\end{figure}

Figure \ref{fig:TBE-cost} (a) and (b) show the letter-value plots for the TBE data (observations) (ranging from 0 to 250 hours) and the TC data (ranging from 0 to 50 million USD). The letter-value plot provides a more accurate visualization compared to the traditional box plots for large and highly skewed data sets (\cite{Hofmann2017}), which is indeed the case in this pipeline accidents dataset. Based on these figures, it's evident that the majority of TBE data points fall within the interval of $(0,25)$ hours, indicating that most failures occur in close succession. Additionally, a significant portion of the TC data is concentrated in the range of 0 to one million dollars, suggesting that most incidents are associated with relatively lower costs. Both these figures suggest right-skewed distributions for both TBE and TC variables while showing a small number of outliers in the dataset.

\section{Risk-adjusted Nonhomogeneous Poisson Process} \label{NHPP}

The NHPP is a counting process that allows the failure rate/intensity to change with time. Denoting an NHPP with $N(t)$ for $t>0$, where $N(t)$ represents the number of failures that occurred in the time interval $(0,t]$. The process $N(t)$ can be fully characterized by the intensity function $\lambda(t)$ that quantifies the failure rate at time $t$. The NHPP  satisfies the following conditions:
\begin{itemize}
\item  $N(0)=0$, which means that no failures occur at time $t=0$.
\item The process has independent increments which means that the number of events in disjoint intervals $\left(s_1, s_2\right]$ and $\left(t_1, t_2\right]$ are independent.
\item At most one failure can occur in any infinitesimal time interval, i.e., $P\left(N(t+\Delta t)-N(t) \geq 2\right)=0$.
\item The probability that a failure occurs in the time interval $[t, t+\Delta t]$ is given by $
 P(N(t+\Delta t)-N(t)=1) = \lambda(t)\Delta t+o(\Delta t)$. 
 \end{itemize}

It follows that in an NHPP, the number of failures in the time interval $(s,t]$ has a Poisson distribution with parameter $\Lambda(t) - \Lambda(s)$ where $\Lambda(t)=E(N(t))=\int_0^t \lambda(u) du$ is called the mean function (also known as cumulative intensity function) of the process. Thus, we have:
\begin{align}\label{NHPP}
 \operatorname{Pr}(N(t)-N(s)=k)=\frac{\left(\Lambda(t)-\Lambda(s)\right)^k}{k !} \exp \left(\Lambda(s)-\Lambda(t)\right), k=0,1,2,...
\end{align}

According to equation \eqref{NHPP}, it can be inferred that $N(t)$ follows a Poisson distribution with the parameter $\Lambda(t)$.  When the intensity function is constant, i.e., $\lambda(t)=\gamma$, say, the NHPP reduces to HPP with the constant rate $\gamma$. The intensity function $\lambda(t)$ can take different mathematical forms. However, in many applications, the log-linear and the power-law models are the most widely used. The following two subsections provide a brief introduction to these models.

\subsection{Power law intensity model}
 The power law intensity function, which is also known as the Weibull model, is a widely used failure rate function in NHPP applications with the mathematical form:
\begin{align}\label{equ:log_power}
\lambda(t) = \gamma \eta t^{\eta-1},
\end{align}
and with the mean function:
\begin{align}\label{equ:Log_power}
 \Lambda(t)=\int_0^{t} \gamma \eta u^{\eta-1}du=\gamma t^{\eta},
 \end{align}
where $\gamma>0$ is the scale and $\eta>0$ is shape parameter. According to \eqref{equ:log_power} and \eqref{equ:Log_power}, $\eta>1$ results in a shorter time between failures, while $0 < \eta < 1$ leads to a longer time between events. If $\eta=1$, the failure process becomes an HPP and the TBE follows an exponential distribution with rate $\gamma$.

\subsection{Log-Linear intensity model}
 A common way to model the intensity function of an NHPP is using a log-linear model with the following mathematical form:
\begin{align}\label{equ:log_log}
\lambda(t) = \exp(\gamma+ \eta t),
\end{align}
where $\gamma$ and $\eta$ are real constants. According to \eqref{equ:log_log}, $\gamma$ corresponds to the initial failure rate and $\eta$ quantifies the magnitude of the failure rate's changes proportional to the time. The mean function corresponding to the log-linear model can also be calculated as:
 \begin{align}\label{equ:Log_log}
 \Lambda(t)=\int_0^{t} \exp(\gamma+\eta u) du=\frac{1}{\eta} \big(\exp(\gamma+\eta t)-\exp(\gamma)\big).
 \end{align}

The values of $\gamma$ and $\eta$ in the above setting allow for modelling a diverse range of time-dependent patterns in failure rates, including accelerating, decelerating, and constant failure rates, over time. The value of $\eta>0$ shows a positive association between the failure intensity and the time, which results in a shorter time between failures, while a negative $\eta$ leads to a longer time between failures. On the other hand, when $\eta=0$, the failure process reduces to an HPP and the TBE follows an exponential distribution with rate $e^\gamma$. Unlike the power law intensity model, the most important characteristic of the log-linear approach is that its failure intensity function is greater than zero at $t=0$ and is convex for any $\eta$ value. This characteristic makes this model suitable to describe the failure process of a repairable system with an extremely fast increasing failure rate.

Within the framework of NHPP, acknowledging the impact of various risk factors on system failure mechanisms emphasizes the essential need for a model that adjusts for these risks. The next subsection discusses the Risk-adjusted NHPP.


%

\subsection{Risk Adjustment }

In many real-world applications, the failure mechanism of a system can be influenced by various risk factors. For instance, in our motivating example, factors such as the pipeline type, the cause of the accident, and the type of hazardous liquid can all affect pipeline failures. When risk factors are present, a risk-adjusted model allows for a separate estimation of the intensity function from the effects of the risk factors. This leads to a more reliable estimate of the system's failure rate.\\

The Cox proportional hazards (PH) model is the most widely employed technique in survival studies for analyzing time-to-event data in the presence of risk factors. This model assumes that risk factors have a multiplicative influence on the hazard function, without necessitating a constant hazard function or adherence to a specific distribution.  The PH model relates the hazard function at time $t$ with the vector of risk factors  $\bm{z}_t=\left(z_{t1}, z_{t2}, \ldots\right.$, $\left.z_{tp}\right)$ collected at this time to a baseline hazard (intensity) function as:
\begin{align}\label{equ:hazard}
\lambda(t| \bm{z}_t)=\lambda(t) \exp \left(\bm{\beta}^{\prime} \bm{z}_t \right), 
\end{align}
where $\bm{\beta}=\left(\beta_1, \beta_2, \ldots, \beta_p\right)$ is a vector of $p$ unknown regression coefficients and $\lambda(t)$ is the baseline intensity function which only depends on time $t$ and is given in equations \eqref{equ:log_power} and \eqref{equ:log_log}. The interpretation of beta values depends on the scale and nature of the risk factors. The risk-adjusted intensity function $\lambda(t| \bm{z}_t)$ in  \eqref{equ:hazard} can be used to derive the risk-adjusted mean function as:
\begin{align}\label{equ:hazardb}
 \Lambda(t|\bm{z}_t)=\int_0^{t} \lambda(u|\bm{z})du=\Lambda(t) \exp \left(\bm{\beta}^{\prime} \bm{z}_t\right),  
\end{align}
where $\Lambda(t)$ is given in equations \eqref{equ:Log_power} and \eqref{equ:Log_log}. As per the equation \eqref{equ:hazard}, the impact of risk factors on failures is quantified through the regression coefficients $\bm{\beta}$. In the presence of one risk factor ($p=1$), a positive $\beta$ value implies that $\exp \left(\beta z_t\right) > 1$, which in turn increases the expected number of failures in the time interval $(0,t]$ and thus reduces the time between failures. Conversely, negative $\beta$ values suggest a decrease in the expected number of failures resulting in a longer time between failures. In the upcoming section, we will delve into the dependence among variables.

\section{Dependence modeling}
 \label{sec:copula}
Copula modelling has emerged as a widely adopted technique for capturing dependencies in various domains of application. It offers the ability to disentangle the dependence structure from the joint distribution function of a set of variables while preserving their individual univariate marginals. Sklar's theorem plays a pivotal role in this process, allowing us to establish the joint distribution $H(x, y)$ of two random variables, namely $X$ and $Y$, with marginal cumulative distribution functions $F_{X}(.|\bm{\theta}_X)$ and $F_{Y}(.|\bm{\theta}_Y)$, respectively. The equation for $H(x, y|\bm{\theta}_X,\bm{\theta}_Y, C)$ involves the copula model $C$ once its parametric form is determined, where:
\begin{align}\label{equ:copula}
H(x,y|\bm{\theta}_X,\bm{\theta}_Y,C)=C\left(F_X(x|\bm{\theta}_X),F_Y(y|\bm{\theta}_Y)|\theta_c\right),
\end{align}
where $\bm{\theta}_X$ ($\bm{\theta}_Y$) is the vector of  $X$ ($Y$) distribution's parameters and $\theta_c$ is the parameter of the copula function $C$. A key advantage of Sklar's theorem lies in its guarantee that a unique copula $C$ exists when dealing with continuous marginal distributions, as is the case in our study. Before delving into specific copula models, it is essential to introduce the dependence measure known as Kendall's tau $\tau$. 

\subsection{Kendall's Tau $\tau$}
Kendall's tau $\tau \in [-1,1]$ is a widely used metric for quantifying dependencies associations between two random variables. For continuous random variables $X$ and $Y$ with a copula $C$, In the context of \citet{Gibbons}, Kendall's tau is defined specifically in terms of cumulative distribution functions (CDFs). This approach highlights the nonparametric nature of Kendall's tau and its independence from specific distributional assumptions. Kendall's tau $\tau$ can be computed using the formula:
\begin{equation}\label{tau_eq}
\tau = 4 \iint_{[0,1]^{2}} C(u, v) \mathrm{d} C(u, v) - 1.
\end{equation}

The coefficient $\tau$ measures the associations between two variables. Values $\tau \in (0,1)$ indicate a positive association between $X$ and $Y$ (both variables increase/decrease together), while a $\tau \in (-1,0)$ expresses a negative association (as one variable increases, the other one decreases). On the other hand, $\tau=1 (\tau=-1)$  shows a perfect positive (negative) association and $\tau=0$  indicates no association between the variables. It is also interesting to note that the integral in \eqref{tau_eq} represents the expected value of the random variable $C(U,V)$, where individual variables $U$ and $V$ follow a standard uniform distribution,i.e., $\tau = 4 \mathbb{E}(C(U, V)) - 1$. Indeed, there exists a diverse range of copula functions with varying properties. In the subsequent subsection, we introduce the Gumbel copula from Archimedean family and highlight its distinctive characteristics.

\subsection{Gumbel copula model}

There could be different dependence structures between the TBE and TC  variables that we aim to monitor jointly over time.  \cite{Castagliola} presented examples in which there is a positive or negative or even a lack of association between TBE and TC variables. However, in the case study they discussed there is a strong positive association which is equivalent to saying that the TC becomes smaller as the TBE becomes shorter. This is also more likely to happen in our case study of monitoring pipeline accidents. Accordingly, this paper focuses on the Gumbel model which belongs to the well-known bivariate Archimedean family of copulas. The Gumbel family can represent only independence and positive dependence, as its dependence parameter is bounded between the independence copula and the Frechet-Hoeffding upper bound copula. It exhibits strong upper tail dependence and relatively weak lower tail dependence. However, the application of the proposed method is not restricted to this choice and other copula functions that are able to model various kinds of dependency can be easily used.

 The Gumbel copula, also known as the Gumbel-Hougaard model, was introduced by \citet{16}. Mathematically, the Gumbel copula is expressed as:
\begin{align}\label{equ:Gumbel}
C(u,v) = \exp\left(-\left(\left(-\ln(u)\right)^{\theta_c} + \left(-\ln(v)\right)^{\theta_c}\right)^{\frac{1}{\theta_c}}\right),
\end{align}
where the dependence parameter $\theta_c$ is confined to the interval $[1, \infty)$. The relationship between Kendall's tau $\tau$ and the Gumbel copula parameter $\theta_c$ is given by:
\begin{align}\label{equ:thetaGumbel}
\tau = \frac{\theta_c-1}{\theta_c} \iff \theta_c = \frac{1}{1-\tau},
\end{align}

 Where the Gumbel copula parameter ($\theta_c$) is utilized within the copula function (Equation \eqref{equ:thetaGumbel}), whereas Kendall's tau ($\tau$) employs to characterize the dependence in numerical analysis in Section \ref{sec:Per.Ev}. 

\section{Proposed dynamic monitoring technique}
 \label{sec:control.chart}

Recall that the main aim of this study is to develop a control chart to monitor the TBE variable and its associated TC while accounting for the risk factors recorded at the time of each failure. In this setting, one reasonable quantity for monitoring the stability of such a process, that incorporates the effects of these two factors, is the average cost per unit time (AC) so that AC=TC/TBE. The average cost per unit time metric also has a meaningful interpretation, making the control chart interpretation more practically meaningful and providing valuable insights into the financial implications. Both of these are relevant aspects of the decision-making process.

\begin{figure}[H]
 \centering
 \includegraphics[scale=0.7]{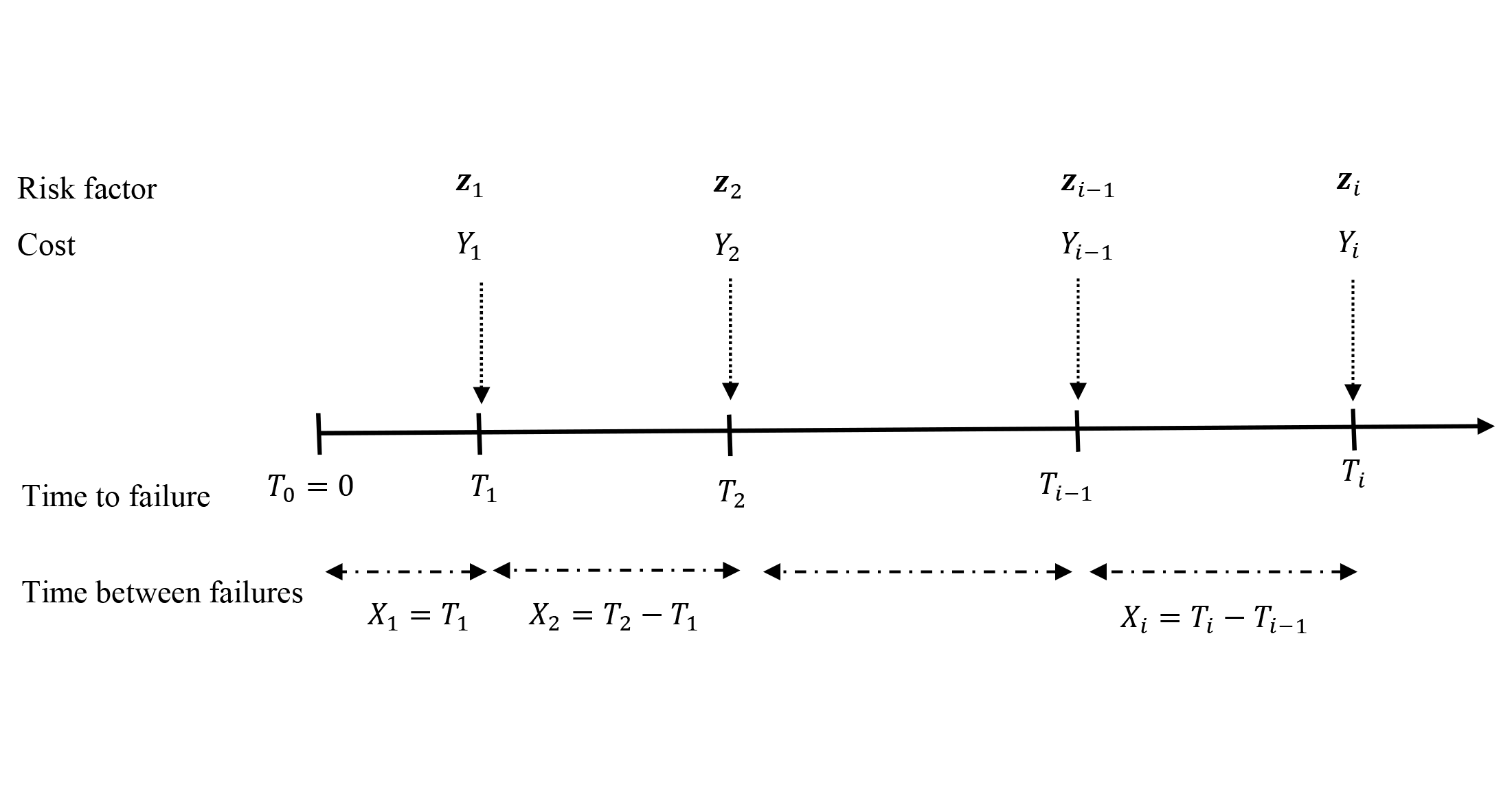} 
 \caption{Data collection schematic.}
 \label{fig:timeline}
 \end{figure}

We assume that the system starts to operate at time $T_0=0$. As time passes, failures may occur at random times $T_1, T_2,...$ resulting in TBEs $X_1=T_1, X_2=T_2-T_1,...$. In addition, at time $t_i$ of failure $i=1,2,\ldots$, data on the risk factors and the TC variable denoted by $\bm{z}_{i1},...,\bm{z}_{ip}$ and  $Y_i$, respectively, are collected. Figure \ref{fig:timeline}  shows a schematic of this data collection process.

To calculate the control limits of the proposed monitoring scheme, we first derive the conditional cumulative distribution function (CDF) of the \(i\)th failure time $T_i$ given the vector of risk factors $\textbf{\textit{z}}_{i}$ and the fact that $T_{i-1}=t_{i-1}$ as:
 \begin{align}\label{equ:distTime}
\notag P\left(T_i \leq t_i \mid T_{i-1}=t_{i-1}, \bm{z_{i}},\bm{\theta}_{X} \right)&=1-P\left(N\left(t_i\right)-N\left(T_{i-1}\right)=0 \mid T_{i-1}=t_{i-1}\right) \\
\notag & =1-P\left(N\left(t_i\right)-N\left(t_{i-1}\right)=0\right)\\
&=1-\exp \left(-\Lambda\left(t_i|\bm{z}_{i}\right)+\Lambda\left(t_{i-1}|\bm{z}_{i}\right)\right),
\end{align}
 where $\bm{\theta}_{X}=(\gamma,\eta,\bm{\beta})$ is the vector of parameters regarding the intensity function and risk model and $\Lambda\left(.|\bm{z}_{i}\right)$ is given in \eqref{equ:hazardb}. The second equality in \eqref{equ:distTime} is obtained using the independent increments of NHPP and the third one is obtained using the fact that:
\begin{align}
\notag P\Big(N(t_{i-1}+x)-&N(t_{i-1})=k|T_{i-1}=t_{i-1}, \bm{z_{i}},\bm{\theta}_{X}\Big)\\
&=\frac{[\Lambda(t_{i-1}+x| \bm{z_{i}})-\Lambda(t_{i-1}|\bm{z}_{i})]^k}{k !} e^{-[\Lambda(t_{i-1}+x| \bm{z_{i}})-\Lambda(t_{i-1}|\bm{z}_{i})]}.
\end{align} 

Eventually, the conditional CDF of $X_i$ given $T_{i-1}=t_{i-1}$ and $\bm{z}_{i}$ can be obtained using equation \eqref{equ:distTime} as follows:
\begin{align}\label{equ:cdf_TBE}
P\left(X_i \leq x|T_{i-1}=t_{i-1},  \bm{z_{i}}, \bm{\theta}_{X}\right)=1-\exp \left(-\Lambda\left(t_{i-1}+x| \bm{z_{i}}\right)+\Lambda\left(t_{i-1}| \bm{z_{i}}\right)\right).
\end{align}
 As a result, the conditional probability density function (PDF) of $X_i$ can be calculated as:
\begin{align}\label{equ:pdf_TBE}
f_{(X_i| t_{i-1}, \bm{z_{i}})}\left(x|\bm{\theta}_{X}\right)= \lambda\left(t_{i-1}+x| \bm{z_{i}}\right) \exp \left(-\Lambda\left(t_{i-1}+x| \bm{z_{i}}\right)+\Lambda\left(t_{i-1}| \bm{z_{i}}\right)\right).
\end{align}

Let $Y_i$ be the corresponding random total cost of failure $i=1,2,\ldots$ having the CDF $F_{Y_i}(.)$ and PDF $f_{Y_i}(.)$ with the vector of parameters $\bm{\theta}_{Y}$. Furthermore, let us assume $X_i$ and $Y_i$ to be dependent and their joint CDF is given by: 
\begin{align}
F_{(X_i, Y_i|t_{i-1}, \bm{z_{i}})}(x,y|\bm{\theta})=C\big( F_{\left(X_i|t_{i-1}, \bm{z_{i}}\right)}(x| \bm{\theta}_{X}), F_{Y_i}(y| \bm{\theta}_{Y})|\theta_c \big),  
\end{align}
 where $C(u,v|\theta_c)$ is a Copula model containing all information on the dependence structure between $Y_i$ and $X_i$ and $\bm{\theta}=\left(\bm{\theta}_{X}, \bm{\theta}_{Y},\theta_c \right)$. 
Accordingly, the joint PDF of $Y_i$ and $X_i$ can be derived as:
 \begin{align}\label{equ:p.d.f}
\notag f_{(X_i, Y_i|t_{i-1}, \bm{z_{i}})}(x,y|\bm{\theta})&=f_{\left(X_i|t_{i-1}, \bm{z_{i}}\right)}(x| \bm{\theta}_{X}). f_{Y_i}(y| \bm{\theta}_{Y})\times \\ 
& c\left( F_{\left(X_i|t_{i-1}, \bm{z_{i}}\right)}(x| \bm{\theta}_{X}), F_{Y_i}(y| \bm{\theta}_{Y})|\theta_c\right),  
\end{align}
where $f_{\left(X_i|t_{i-1}, \bm{z_{i}}\right)}(.| \bm{\theta}_{X})$ is given in \eqref{equ:pdf_TBE}, $f_{Y_i}(.| \bm{\theta}_{Y})$ is the PDF of $Y_i$, and $c(u,v\mid \theta_c)=\frac{\partial C(u,v\mid \theta_c)}{\partial u \partial v}$ is the copula density.

 \subsection{The monitoring statistic}
 According to the provided background in the previous subsection, the average cost per time unit at the time of failure $i$ could be mathematically defined by $AC=E\left(\frac{Y_i}{X_i}\right)$ and be calculated as:
\begin{align}
E\left( \frac{Y_i}{X_i}| t_{i-1}, \bm{z_{i}},\bm{\theta} \right)= \int_{0}^{\infty} \int_{0}^{\infty} \frac{y_i}{x_i} f_{\left(X_i, Y_i|t_{i-1}, \bm{z_{i}}, \bm{\theta} \right)}(x,y) dy_i dx_i,
\end{align}
where $f_{(X_i, Y_i|t_{i-1}, \bm{z_{i}},\bm{\theta})}$ is given in \eqref{equ:p.d.f}. By the definition, a small value of AC is typically desired as it indicates lower average costs incurred per failure.  We propose to calculate the following monitoring statistic at the time of failure $i$:
\begin{align}
W_i = \frac{Y_i}{X_i}, \qquad \text{for} \qquad i=1,2,\ldots,
\end{align}
 with the aim of detecting possible deviations of the AC from its in-control (IC) state to an out-of-control (OC) state as quickly as possible. Let $\bm{\theta}_{0}$ and $\bm{\theta}_{1}$ denote the vectors of IC and OC  parameters. Accordingly, monitoring the AC over time could be done by developing a proper control chart for the sequence of $W_i$. The proposed monitor statistic $W_i$ further satisfies the following properties: 
\begin{itemize}
\item[(1)] $W_i$ decreases with $X_i$ and increases with $Y_i$.
\item[(2)] $W_i$ increases with $X_i$ and decreases with $Y_i$.
\end{itemize}

Let $F_{W_i}(.|t_{i-1}, \bm{z_{i}},\bm{\theta})$ be the conditional CDF of $W_i$ given $t_{i-1}$, $ \bm{z_{i}}$ and $\bm{\theta}$. 
 Then, the CDF of the monitoring statistic $W_i$ can be calculated as follows:
\begin{align}\label{equ:Wdist}
\notag F_{W_i}(w|t_{i-1}, \bm{z_{i}},\bm{\theta})&=P\left(W_i\leq w| t_{i-1}, \bm{z_{i}},\bm{\theta}\right)\\
\notag &=P\left(   \frac{Y_i}{X_i}\leq w|t_{i-1}, \bm{z_{i}},\bm{\theta}\right)\\
  &= \int_0^\infty \int_0^{xw} f_{(X_i, Y_i|t_{i-1}, \bm{z_{i}})}(x,y|\bm{\theta}) dy dx,
\end{align}
where $f_{(X_i, Y_i|t_{i-1}, \bm{z_{i}})}(.,.|\bm{\theta})$ is given in \eqref{equ:p.d.f}.

While a higher shift in the AC may be more of interest to detect from a financial point of view, detecting lower shifts can be valuable in assessing any improvements made to the process. Therefore, we propose implementing a two-sided control chart capable of detecting changes in the AC that may occur in either direction, above or below the IC value. The probability-type control limits for the proposed chart at the time of failure $i=1,2,\ldots$ can be calculated as:
\begin{align}
L C L_i &=F_{W_i}^{-1}\left(\frac{\alpha}{2} \mid t_{i-1}, \bm{z_{i}},\bm{\theta}_0 \right),\nonumber\\
U C L_i &=F_{W_i}^{-1}\left(1-\frac{\alpha}{2} \mid t_{i-1}, \bm{z_{i}},\bm{\theta}_0 \right),\label{equ:cl}
\end{align}
where  $\alpha $ is the probability of a type I error and $F_{W_i}^{-1}\left(\ldots \mid  t_{i-1}, \bm{z_{i}},\bm{\theta}_0 \right)$ is the inverse of the CDF of $W_i$ given in \eqref{equ:Wdist}. Thus the control limits can be obtained by solving equations $F_{W_i}(w|t_{i-1}, \bm{z_{i}},\bm{\theta}_0) =\frac{\alpha}{2}$ and $F_{W_i}(w|t_{i-1}, \bm{z_{i}},\bm{\theta}_0)=1- \frac{\alpha}{2}$ for $w$ numerically. Note that the index $i$ of $LCL_i$ and $UCL_i$ in \eqref{equ:cl} shows that they are step-wise limits and should be calculated at the time of each failure based on the information from the previous failure ($t_{i-1}$) and current failure ($\bm{z}_{i}$) as well as the IC vector $\bm{\theta}_0$. Thus, at the time of failure $i=1,2,\ldots$, one needs to first calculate the control limits and then compare $W_i$ with them; if $W_i >UCL_i$ or $W_i<LCL_i$, the control chart triggers an OC signal, if not, process monitoring continues.\\

Average run length (ARL) is one of the most popular measures to assess the performance of a control chart and to compare it with alternative charts. $ARL$ is the expected value of Run Length (RL), defined as the average number of samples taken before the first OC  signal (the monitoring statistic falls beyond the control limits). When the successive monitoring statistics are independent, the IC RL (RL$_0$) follows a Geometric distribution with probability $\alpha$ where $\alpha$ is the probability that an IC process is diagnosed as OC by the control chart. In this case, we have $ARL_0=\frac{1}{\alpha}$. it is shown that the in-control run length RL$_0$ of the proposed control chart is also distributed as a Geometric random variable. We have, when the process is in-control,
\begin{align}\label{equ:arl0}
\notag P(RL_0>n)&=P\left( \bigcap_{i=1}^n \{ LCL_i<W_i<UCL_i\}\right)\\&
\notag =\prod_{i=1}^n P\left( LCL_i<W_i<UCL_i\mid t_{i-1}, \bm{z_{i}},\bm{\theta}_0 \right)\\&
\notag =\prod_{i=1}^n\left(1-\alpha\right)\\&
=\left(1-\alpha\right)^n,
\end{align}
where the second equality is obtained due to the independence between $W_i$ for $i=1,2,\ldots$ which follows from the independent-increments feature of the NHPP. The OC ARL ($ARL_1$) of the proposed chart can be calculated using the Monte-Carlo approach which is described in Section \ref{sec:Per.Ev}.

  \section{Numerical Analysis} 
  \label{sec:Per.Ev}
 
In order to assess the performance of the proposed control chart and investigate its sensitivity with respect to various factors, this section aims to conduct a numerical study based on the $ARL$ metric. First of all, let us set  $ARL_0=200$. In addition, consider a system where its failures come from an NHPP  with power law or log-linear intensity function. The IC parameters are assumed to be $\eta_0=1.50$ for the power law model,  $\eta_0=2.00$ for the log-linear model, and $\gamma_0=0.05$. These choices of intensity parameters represent increasing failure rates, a scenario more probable in deteriorating systems. Furthermore, we assume the associated TC represented by $Y$ follows an Exponential distribution with IC rate $\mu_0=1.00$. The joint distribution of $Y$ and $X$ is also characterized by a Gumbel copula model. Furthermore, we introduce a risk factor $z$ ($p=1$), which impacts the TBE. The risk factor will be a factor ranging from 1 to 3, generated using a truncated Poisson distribution with rate $1$, which means the numbers 1, 2 and 3 will be generated with probabilities $0.36, 0.18$ and  $0.06$, respectively.  In addition, the regression coefficient $\beta$ takes values from the set $\{-2.00,-0.50,0.00,0.50,2.00\}$ to cover different association levels between the risk factor and TBE. 
%
To capture different levels of dependence between the TBE and the TC when they are positively associated, we consider the dependence parameter $\tau$ from the set $\{0.3, 0.8\}$. This will enable us to explore moderate and strong positive dependencies. \\

To conduct the numerical analysis, we assume the parameters $\gamma$ and $\eta$ regarding the TBE variable and the parameter $\mu$ of the TC variable are subject to shift, while $\beta$ and $\theta_c/\tau$ are assumed to remain unchanged. Let us donete the vector of IC parameters by $\bm{\theta}_0=(\gamma_0,\eta_0,\mu_0)$. To calculate the $ARL_1$ values, the vector of OC parameters is also defined as $\bm{\theta}_1=(\gamma_1,\eta_1,\mu_1)=(\delta_{\gamma} \gamma_0,  \delta_{\eta} \eta_0, \delta_{\mu} \mu_0)$ where $\delta>0$ quantifies the shift magnitude in each parameter. We set $\delta_k=0.25, 0.75,1.25, 2.00$ for $k=\gamma, \eta$ and $\mu$ to account for both upward and downward shifts in the associated parameters. By examining the $ARL_1$ values in these cases, we can gain insights into the performance and effectiveness of the control charts under various scenarios and deviations from the stable condition.\\

 $ARL$ values are computed through Monte Carlo simulations described in  \textbf{Algorithm 1}. Though $ARL_0$ can be calculated theoretically by $ARL_0=\frac{1}{\alpha}$ from \eqref{equ:arl0}, it could also be calculated from \textbf{Algorithm 1} by generating $X_i$ and $Y_i$ from $\bm{\theta}_0$. Moreover, $ARL_1$ can be computed from \textbf{Algorithm 1} by generating $X_i$ and $Y_i$ using $\bm{\theta}_1$. Tables  \ref{tab:pow2} and \ref{tab:log2} show the results of the ARL analysis for power-law and log-linear intensity models, respectively. The primary result shown by the tables is that the proposed chart is $ARL$-unbiased across all shift combinations and both intensity models, i.e.,  we always have $ARL_0>ARL_1$. \\

\begin{algorithm}[H]
\caption{ARL calculation algorithm.}\label{algho-ATS}
\SetKwInOut{Input}{Input}
\SetKwInOut{Output}{Output}
\Input{ $\alpha, \gamma, \eta,\mu, \theta_c, \beta, \tau$.}
\Output{ARL}
RL $\leftarrow$ (...) \Comment{Initialize the vector to store RL values}\\
\For{i=1 to large number}{
    j $\leftarrow 1;  t_0 \leftarrow 0$; \Comment{Time to start ($t_0 = 0$).}\\
    \Repeat{$W_j > UCL_j$ or $W_j < LCL_j$}{
        \For{k=1 to N}{
            $(U_k, V_k) \sim C(U, V | \theta_c$);\\
            $Xs_{k} \leftarrow F^{-1}_X(U_{k} | t_{j-1}, \bm{z}_j, \bm{\theta}_{X})$, $Ys_{k} \leftarrow F^{-1}_Y(V_{k} | \mu);$\\
            $Ws_{k} \leftarrow \frac{Ys_{k}}{Xs_{k}}$;
        }
       $ UCL_j \leftarrow Quantile(Ws, 1-\frac{\alpha}{2})$;\\
        $LCL_j \leftarrow Quantile(Ws, \frac{\alpha}{2})$;\\
     $(U, V) \sim C(U, V | \theta_c)$; \Comment{For next failure};\\
      Generate $\bm{z}_{j}$;\\
        $X_j \leftarrow F^{-1}_{X_j}(U | t_{k-1}, \bm{z}_j, \bm{\theta}_{X}), Y_j \leftarrow F^{-1}_Y(V | \mu), W_j \leftarrow \frac{Y_j}{X_j}$; \label{line}\\
        $t_j \leftarrow t_{j-1} + X_j$;\\
        j $\leftarrow$ j+1;
    }
    RL[i] $\leftarrow$ j;
}
ARL $\leftarrow$  E(RL)
\end{algorithm}

\subsection{Sensitivity analysis}

Let us start by assessing the ability of the proposed chart to detect shifts in  $\gamma$. The tables show that when $\delta_\eta <1$, the chart usually detects downward shifts in $\gamma$ sooner than its upward shifts. For example in the power law intensity function, when $\beta=-2.00, \tau=0.3, \delta_{\eta}=0.75$ and $\delta_{\mu}=0.25$, the $ARL_1$ corresponding to $\delta_{\gamma}=0.25, 0.75, 1.25, 2.00$ are 19.7, 55.7, 88.6 and 128.3, respectively. On the contrary, when $\delta_\eta >1$, the control chart detects upward shifts in $\gamma$ sooner than its downward shifts. For example, in the log-linear intensity function, when $\beta=0.00, \tau=0.8, \delta_{\eta}=2.00$ and $\delta_{\mu}=2.00$, the $ARL_1$ corresponding to $\delta_{\gamma}=0.25, 0.75, 1.25, 2.00$ are 51.9, 41.0, 31.5  and 20.5, respectively. Assessing the chart's performance in detecting shifts in $\eta$ is another important case due to its impact on system deterioration. An increase in the shape parameter $\eta$ leads to an increase in the frequency of failures. We are also interested in evaluating the performance of the chart when a shift has occurred in the mean of TC $\mu$. The observed trend in the chart's ability to identify shifts in $\eta$ appears consistent with its performance in detecting shifts in $\gamma$. This suggests that the chart usually detects downward (upward) shifts in $\eta$ and $\mu$ sooner than its upward (downward) deviations when $\delta_\eta <1$ ($\delta_\eta >1$).  For example in the power law intensity function, when $\beta=2.00, \tau=0.3, \delta_{\gamma}=0.25$ and $\delta_{\mu}=1$, the $ARL_1$ corresponding to $\delta_{\eta}=0.25, 0.75, 1.25, 2.00$ are 4.1, 86.7, 129.4  and 49.8, respectively. We are also interested in evaluating the performance of the chart when a shift has occurred in the mean of TC $\mu$. As an example regarding shifts in $\mu$, in the case of the power law intensity function when $\beta=0.00, \tau=0.3, \delta_{\gamma}=1.25$, and $\delta_{\eta}=2.00$, the $ARL_1$ corresponding to $\delta_{\mu}=0.25, 0.75, 1.00, 1.25, 2.00$ are 13.5, 26.9, 33.2, 38.9, and 52.2 respectively. The reason for these phenomena is that the sampling distribution of the estimators is right-skewed with a longer tail on the right than on the left.\\

It is well known that the risk factors impact the failure mechanism of the system. On the other hand, we know that $\beta=0$ suggests no effect of the risk factor on the hazard rate. In this case, the chart's sensitivity to shifts remains at a baseline level, not influenced by the risk factor.
 By looking at the tables, it can be observed that the control chart declares an OC alarm when $\beta<0 \, (\beta>0)$ sooner (later) than in the scenario of the absence of a risk factor, denoted by $\beta=0$. For example, in the log-linear intensity function, when $\tau=0.3, \delta_{\gamma}=0.25, \delta_{\eta}=2.00$ and $\delta_{\mu}=1.00$, the $ARL_1$ corresponding to $\beta=-2.00,-0.50, 0.50, 2.00$ are 36.0, 48.1, 62.6, 78.4, and 128.9, respectively. This indicates that risk factors exhibiting a negative association with TBE have a positive impact on the performance of the proposed chart, while those with a positive association negatively impact the efficiency of the monitoring technique. This insight stands out as a key finding in our simulation study. This trend could be justified by understanding how the sign of $\beta$ relates to the sensitivity of the control chart to detect shifts. A $\beta<0$ suggests a reduction in the hazard rate associated with the risk factor implying that the effect of the risk factor is associated with a lower likelihood of the shift occurring. Consequently, the chart is more sensitive to changes, detecting shifts sooner because the risk factor is acting as a mitigating or stabilizing factor. On the other hand, a $\beta>0$  implies an increase in the hazard rate associated with the risk factor suggesting that the risk factor is associated with an elevated likelihood of the shift occurring. Thus, the chart is less sensitive to changes, detecting shifts later, as the risk factor contributes to a higher baseline risk of shifts.\\

It can also be observed from the tables that the dependency parameter $\tau$ noticeably affects the OC performance of the proposed monitoring scheme. From the tables, $ARL_1$ decreases when $\tau$ increases. This means that the chart detects the OC conditions sooner when the degree of dependency between TBE and TC is stronger. Thus, the higher degree of dependency between TBE and TC improves the performance of the chart. This finding is also a crucial outcome of our simulation study. This could be justified by the fact that when there is a higher degree of dependency between TBE and TC, it implies that they are more closely related or influenced by each other. As a result, changes in one variable are likely to be reflected in the other and consequently, the chart becomes more sensitive to changes in the joint distribution of the two variables. \\

\begin{sidewaystable}[!htp]
  \centering
  \small 
 \setlength{\tabcolsep}{1.2pt}
\renewcommand{\arraystretch}{1}
   \caption{The $ARL$ values based on power law intensity with $\gamma_0=0.05, \eta_0=1.50, \mu_0=1.00$ when $\tau=0.3$ (upper block) and $\tau=0.8$ (lower block).}
  \begin{tabular}{ccccccccccccccccccccccccccccccc}
    \hline \hline
     &  & \multicolumn{5}{c}{$\beta=-2.00$}&& \multicolumn{5}{c}{$\beta=-0.50$}&& \multicolumn{5}{c}{$\beta=0.00$}&& \multicolumn{5}{c}{$\beta=0.50$}&& \multicolumn{5}{c}{$\beta=2.00$}\\
    \cline{3-7} \cline{9-13}\cline{15-19}\cline{21-25}\cline{27-31}
      $(\delta_{\gamma}, \delta_{\eta})$ & $\delta_{\mu}$ & 0.25 & 0.75 &1.00& 1.25 & 2.00 & ~ & 0.25 & 0.75 &1.00& 1.25 & 2.00 & ~&0.25 & 0.75 &1.00& 1.25 & 2.00 & ~ & 0.25 & 0.75 &1.00& 1.25 & 2.00& ~ & 0.25 & 0.75 &1.00&1.25 & 2.00 \\ \hline
     (0.25,0.25) & ~ & 1.3 & 1.2 & 1.2 & 1.1 & 1.1 & ~ & 1.9 & 1.7 & 1.7 & 1.6 & 1.5 & ~ & 2.4 & 2.1 & 2.1 & 2.0 & 1.9 & ~ & 2.9 & 2.7 & 2.6 & 2.6 & 2.4 & ~ & 4.1 & 4.0 & 3.9 & 3.8 & 3.9 \\ 
        (0.75,0.25) & ~ & 1.7 & 1.5 & 1.5 & 1.5 & 1.4 & ~ & 3.0 & 2.7 & 2.6 & 2.5 & 2.4 & ~ & 3.9 & 3.5 & 3.4 & 3.3 & 3.1 & ~ & 4.8 & 4.5 & 4.4 & 4.2 & 4.2 & ~ & 3.5 & 4.7 & 5.1 & 5.0 & 5.1 \\ 
        (1.00,1.00) & ~ & 80.9 & 180.8 & 198.9 & 181.3 & 152.7 & ~ & 83.0 & 188.7 & 196.2 & 195.1 & 143.7 & ~ & 81.9 & 182.2 & 196.7 & 192.5 & 144.0 & ~ & 82.8 & 174.1 & 198.6 & 187.6 & 148.1 & ~ & 83.6 & 178.9 & 197.7 & 176.2 & 143.8 \\ 
        (1.25,0.25) & ~ & 1.9 & 1.7 & 1.7 & 1.7 & 1.6 & ~ & 3.7 & 3.3 & 3.2 & 3.2 & 3.0 & ~ & 4.8 & 4.5 & 4.3 & 4.2 & 4.1 & ~ & 5.1 & 5.3 & 5.3 & 5.4 & 5.2 & ~ & 1.9 & 3.3 & 3.7 & 4.2 & 4.8 \\ 
        (2.00,0.25) & ~ & 2.2 & 2.0 & 2.0 & 1.9 & 1.9 & ~ & 4.6 & 4.2 & 4.0 & 4.0 & 3.8 & ~ & 5.0 & 5.5 & 5.3 & 5.2 & 5.0 & ~ & 4.2 & 5.6 & 5.9 & 6.0 & 6.1 & ~ & 1.1 & 1.7 & 2.0 & 2.3 & 3.3 \\ 
        (0.25,0.75) & ~ & 19.7 & 9.3 & 7.7 & 6.8 & 5.2 & ~ & 41.1 & 18.5 & 14.6 & 12.2 & 9.2 & ~ & 50.7 & 22.8 & 17.9 & 15.2 & 11.0 & ~ & 59.2 & 26.1 & 21.8 & 18.3 & 12.7 & ~ & 86.7 & 40.7 & 33.0 & 27.6 & 19.4 \\ 
        (0.75,0.75) & ~ & 55.7 & 25.9 & 20.9 & 18.0 & 12.8 & ~ & 114.7 & 55.3 & 44.4 & 37.7 & 26.1 & ~ & 132.7 & 64.6 & 53.3 & 45.5 & 32.0 & ~ & 147.2 & 76.3 & 62.9 & 53.0 & 38.2 & ~ & 152.0 & 111.1 & 91.4 & 80.4 & 57.4 \\ 
        (1.25,0.75) & ~ & 88.6 & 41.7 & 33.5 & 28.4 & 20.7 & ~ & 149.6 & 91.5 & 70.3 & 61.4 & 44.5 & ~ & 161.0 & 106.0 & 85.9 & 76.8 & 53.3 & ~ & 155.5 & 124.2 & 98.1 & 91.7 & 63.8 & ~ & 97.3 & 152.3 & 134.4 & 121.9 & 93.5 \\ 
        (2.00,0.75) & ~ & 128.3 & 64.1 & 53.1 & 44.8 & 32.2 & ~ & 160.2 & 127.1 & 108.9 & 93.0 & 69.6 & ~ & 128.9 & 150.7 & 125.1 & 113.5 & 83.4 & ~ & 103.8 & 163.2 & 143.0 & 128.8 & 97.7 & ~ & 44.1 & 154.7 & 162.4 & 157.9 & 130.3 \\ 
        (0.25,1.25) & ~ & 47.1 & 110.3 & 138.0 & 159.6 & 179.3 & ~ & 81.2 & 163.3 & 180.4 & 186.3 & 159.5 & ~ & 91.6 & 177.8 & 185.5 & 181.0 & 148.0 & ~ & 99.0 & 181.0 & 180.3 & 172.6 & 134.0 & ~ & 129.4 & 174.0 & 158.6 & 140.1 & 82.5 \\ 
        (0.75,1.25) & ~ & 22.1 & 58.1 & 70.7 & 90.7 & 120.7 & ~ & 38.5 & 93.4 & 118.4 & 138.4 & 169.9 & ~ & 45.1 & 105.3 & 129.1 & 146.8 & 172.6 & ~ & 49.6 & 116.6 & 144.1 & 159.3 & 187.7 & ~ & 64.0 & 143.0 & 168.1 & 181.4 & 169.1 \\ 
        (1.25,1.25) & ~ & 15.7 & 40.9 & 51.6 & 60.1 & 91.7 & ~ & 26.7 & 68.3 & 87.9 & 104.7 & 144.4 & ~ & 30.8 & 76.0 & 97.2 & 115.0 & 157.4 & ~ & 35.6 & 88.6 & 108.7 & 123.8 & 167.9 & ~ & 45.6 & 105.5 & 134.1 & 147.8 & 182.5 \\ 
        (2.00,1.25) & ~ & 11.6 & 29.7 & 37.1 & 44.2 & 67.3 & ~ & 19.6 & 50.7 & 63.1 & 75.9 & 108.4 & ~ & 21.7 & 58.1 & 72.5 & 89.7 & 122.4 & ~ & 25.2 & 64.2 & 81.2 & 97.0 & 132.0 & ~ & 32.9 & 80.2 & 103.8 & 123.3 & 160.8 \\ 
        (0.25,2.00) & ~ & 6.7 & 13.4 & 16.2 & 18.4 & 24.5 & ~ & 17.1 & 35.0 & 43.8 & 50.2 & 68.3 & ~ & 22.5 & 47.2 & 57.8 & 65.6 & 89.3 & ~ & 28.4 & 60.1 & 70.3 & 79.1 & 106.9 & ~ & 49.8 & 96.5 & 112.4 & 121.7 & 139.7 \\ 
        (0.75,2.00) & ~ & 4.8 & 9.5 & 11.4 & 12.9 & 17.3 & ~ & 12.0 & 24.8 & 29.7 & 35.1 & 47.8 & ~ & 15.6 & 31.8 & 38.8 & 45.1 & 61.7 & ~ & 19.7 & 41.5 & 49.2 & 56.5 & 75.7 & ~ & 34.5 & 71.6 & 82.8 & 97.0 & 116.4 \\ 
        (1.25,2.00) & ~ & 4.2 & 8.2 & 9.8 & 11.4 & 15.1 & ~ & 10.1 & 21.1 & 25.8 & 29.4 & 39.8 & ~ & 13.5 & 26.9 & 33.2 & 38.9 & 52.2 & ~ & 16.5 & 33.9 & 41.9 & 48.4 & 66.1 & ~ & 29.0 & 60.6 & 71.4 & 83.7 & 107.2 \\ 
        (2.00,2.00) & ~ & 3.7 & 7.2 & 8.5 & 9.8 & 12.5 & ~ & 8.8 & 18.1 & 21.7 & 24.9 & 34.4 & ~ & 11.2 & 23.7 & 28.0 & 33.6 & 46.0 & ~ & 14.2 & 30.4 & 35.8 & 41.5 & 57.4 & ~ & 24.8 & 52.7 & 64.2 & 69.0 & 92.2 \\ 
      \hline
        (0.25,0.25) & ~ & 1.1 & 1.0 & 1.0 & 1.0 & 1.0 & ~ & 1.4 & 1.3 & 1.3 & 1.2 & 1.2 & ~ & 1.7 & 1.5 & 1.4 & 1.4 & 1.3 & ~ & 1.8 & 1.8 & 1.7 & 1.7 & 1.6 & ~ & 1.3 & 1.6 & 1.6 & 1.6 & 1.7 \\ 
        (0.75,0.25) & ~ & 1.3 & 1.2 & 1.2 & 1.2 & 1.1 & ~ & 1.8 & 1.8 & 1.7 & 1.7 & 1.6 & ~ & 1.7 & 1.9 & 1.9 & 1.9 & 1.9 & ~ & 1.3 & 1.7 & 1.8 & 1.8 & 1.9 & ~ & 1.0 & 1.0 & 1.1 & 1.1 & 1.2 \\ 
        (1.00,1.00) & ~ & 5.5 & 159.4 & 197.9 & 171.2 & 95.7 & ~ & 6.0 & 161.9 & 195.0 & 174.3 & 94.5 & ~ & 6.0 & 163.7 & 197.6 & 180.7 & 90.7 & ~ & 6.3 & 156.8 & 198.8 & 178.5 & 93.2 & ~ & 5.6 & 158.3 & 201.5 & 176.5 & 86.3 \\ 
        (1.25,0.25) & ~ & 1.5 & 1.4 & 1.3 & 1.3 & 1.2 & ~ & 1.7 & 1.9 & 1.9 & 1.9 & 1.9 & ~ & 1.3 & 1.8 & 1.7 & 1.9 & 2.0 & ~ & 1.1 & 1.3 & 1.4 & 1.5 & 1.6 & ~ & 1.3 & 1.0 & 1.0 & 1.0 & 1.0 \\ 
        (2.00,0.25) & ~ & 1.5 & 1.5 & 1.5 & 1.4 & 1.4 & ~ & 1.4 & 1.8 & 1.8 & 1.9 & 2.0 & ~ & 1.0 & 1.4 & 1.4 & 1.6 & 1.8 & ~ & 1.0 & 1.1 & 1.1 & 1.1 & 1.3 & ~ & 1.1 & 1.7 & 1.1 & 1.0 & 1.0 \\ 
        (0.25,0.75) & ~ & 4.2 & 2.4 & 2.2 & 2.0 & 1.8 & ~ & 10.9 & 3.2 & 2.7 & 2.4 & 2.1 & ~ & 15.3 & 4.0 & 3.2 & 2.7 & 2.2 & ~ & 20.5 & 5.2 & 3.9 & 3.2 & 2.4 & ~ & 30.0 & 10.6 & 7.3 & 5.8 & 3.6 \\ 
        (0.75,0.75) & ~ & 17.6 & 5.8 & 4.4 & 3.7 & 2.8 & ~ & 38.1 & 18.1 & 12.3 & 9.3 & 5.1 & ~ & 29.3 & 24.7 & 17.3 & 13.1 & 6.9 & ~ & 25.5 & 33.3 & 23.6 & 17.7 & 9.6 & ~ & 7.6 & 48.8 & 42.6 & 34.6 & 19.6 \\ 
        (1.25,0.75) & ~ & 23.8 & 12.0 & 8.5 & 6.7 & 4.4 & ~ & 21.5 & 41.4 & 28.7 & 22.3 & 11.8 & ~ & 9.7 & 55.3 & 41.3 & 30.9 & 17.1 & ~ & 4.1 & 66.5 & 52.7 & 41.9 & 23.4 & ~ & 1.0 & 46.2 & 60.8 & 63.1 & 44.0 \\ 
        (2.00,0.75) & ~ & 22.9 & 24.3 & 17.7 & 13.9 & 7.8 & ~ & 3.0 & 70.4 & 59.6 & 47.5 & 26.9 & ~ & 1.2 & 65.3 & 76.4 & 63.0 & 36.8 & ~ & 1.0 & 52.4 & 74.5 & 78.2 & 50.3 & ~ & 1.0 & 8.9 & 31.3 & 61.5 & 74.7 \\ 
        (0.25,1.25) & ~ & 6.3 & 56.8 & 86.9 & 107.3 & 155.1 & ~ & 23.6 & 128.5 & 158.1 & 164.9 & 95.2 & ~ & 32.8 & 147.3 & 162.7 & 157.5 & 60.3 & ~ & 42.1 & 166.6 & 164.6 & 132.4 & 36.9 & ~ & 70.3 & 142.5 & 93.1 & 54.4 & 8.5 \\ 
        (0.75,1.25) & ~ & 1.5 & 14.3 & 23.8 & 35.0 & 67.1 & ~ & 3.1 & 41.3 & 61.7 & 82.2 & 144.5 & ~ & 4.5 & 52.1 & 80.4 & 103.9 & 153.2 & ~ & 6.3 & 64.4 & 95.2 & 124.7 & 168.5 & ~ & 14.3 & 96.9 & 133.7 & 156.1 & 124.5 \\ 
        (1.25,1.25) & ~ & 1.2 & 6.0 & 11.9 & 17.9 & 37.7 & ~ & 1.6 & 20.4 & 34.7 & 49.5 & 92.1 & ~ & 1.8 & 27.3 & 43.3 & 60.9 & 110.9 & ~ & 2.4 & 35.1 & 55.0 & 73.5 & 129.1 & ~ & 5.2 & 57.8 & 86.0 & 109.5 & 157.9 \\ 
        (2.00,1.25) & ~ & 1.1 & 2.9 & 5.4 & 8.8 & 21.2 & ~ & 1.2 & 9.3 & 17.7 & 26.7 & 55.2 & ~ & 1.3 & 12.7 & 23.4 & 34.5 & 69.3 & ~ & 1.4 & 17.6 & 30.3 & 42.8 & 81.1 & ~ & 2.4 & 32.1 & 49.2 & 69.4 & 120.3 \\ 
        (0.25,2.00) & ~ & 1.4 & 2.8 & 3.6 & 4.4 & 6.7 & ~ & 2.5 & 10.3 & 14.3 & 17.7 & 29.5 & ~ & 3.8 & 16.3 & 21.9 & 27.6 & 42.2 & ~ & 6.0 & 23.4 & 31.3 & 39.3 & 57.6 & ~ & 17.9 & 49.5 & 57.7 & 63.4 & 50.3 \\ 
        (0.75,2.00) & ~ & 1.2 & 1.9 & 2.4 & 2.9 & 4.2 & ~ & 1.6 & 5.4 & 7.7 & 9.9 & 16.6 & ~ & 2.1 & 8.7 & 12.0 & 15.6 & 25.1 & ~ & 2.9 & 13.0 & 17.6 & 22.5 & 35.3 & ~ & 9.4 & 31.0 & 39.7 & 47.5 & 59.3 \\ 
        (1.25,2.00) & ~ & 1.1 & 1.7 & 2.0 & 2.4 & 3.5 & ~ & 1.4 & 3.9 & 5.6 & 7.4 & 12.6 & ~ & 1.7 & 6.2 & 9.0 & 11.9 & 19.1 & ~ & 2.3 & 9.6 & 13.3 & 17.2 & 27.8 & ~ & 6.8 & 24.4 & 31.9 & 37.9 & 52.5 \\ 
        (2.00,2.00) & ~ & 1.1 & 1.5 & 1.8 & 2.1 & 2.9 & ~ & 1.3 & 3.1 & 4.3 & 5.7 & 9.6 & ~ & 1.5 & 4.7 & 7.0 & 9.0 & 14.9 & ~ & 1.8 & 7.3 & 10.4 & 13.3 & 21.8 & ~ & 5.2 & 18.7 & 25.5 & 30.9 & 43.9 \\ \hline \hline 
   \end{tabular} \label{tab:pow2}
\end{sidewaystable}

\begin{sidewaystable}[!htp]
  \centering
  \small
 \setlength{\tabcolsep}{1.2pt}
\renewcommand{\arraystretch}{1}
   \caption{The $ARL$ values based on log-linear intensity with $\gamma_0=0.05, \eta_0=2.00, \mu_0=1.00$ when $\tau=0.3$ (upper block) and $\tau=0.8$ (lower block).}
  \begin{tabular}{ccccccccccccccccccccccccccccccc}
    \hline \hline
     &  & \multicolumn{5}{c}{$\beta=-2.00$}&& \multicolumn{5}{c}{$\beta=-0.50$}&& \multicolumn{5}{c}{$\beta=0.00$}&& \multicolumn{5}{c}{$\beta=0.50$}&& \multicolumn{5}{c}{$\beta=2.00$}\\
    \cline{3-7} \cline{9-13}\cline{15-19}\cline{21-25}\cline{27-31}
      $(\delta_{\gamma}, \delta_{\eta})$ & $\delta_{\mu}$ & 0.25 & 0.75 &1.00& 1.25 & 2.00 & ~ & 0.25 & 0.75 &1.00& 1.25 & 2.00 & ~&0.25 & 0.75 &1.00& 1.25 & 2.00 & ~ & 0.25 & 0.75 &1.00& 1.25 & 2.00& ~ & 0.25 & 0.75 &1.00&1.25 & 2.00 \\ \hline
    (0.25,0.25) & ~ & 3.4 & 3.3 & 3.2 & 3.1 & 2.5 & ~ & 7.1 & 5.5 & 5.2 & 4.9 & 4.4 & ~ & 11.3 & 8.2 & 7.7 & 7.1 & 6.2 & ~ & 17.4 & 12.6 & 11.3 & 10.6 & 9.0 & ~ & 51.7 & 34.7 & 30.2 & 26.8 & 22.0 \\ 
        (0.75,0.25) & ~ & 3.8 & 3.9 & 3.8 & 3.8 & 2.9 & ~ & 10.6 & 8.1 & 7.4 & 7.0 & 6.1 & ~ & 17.7 & 12.9 & 11.8 & 11.1 & 9.3 & ~ & 27.9 & 19.8 & 17.8 & 16.9 & 14.0 & ~ & 77.0 & 54.2 & 47.8 & 43.7 & 35.6 \\ 
        (1.00,1.00) & ~ & 56.9 & 162.5 & 199.0 & 206.5 & 132.6 & ~ & 81.4 & 179.2 & 198.1 & 184.2 & 144.1 & ~ & 79.1 & 186.6 & 196.9 & 183.4 & 142.2 & ~ & 80.9 & 176.2 & 195.8 & 182.1 & 146.0 & ~ & 84.3 & 180.3 & 199.0 & 194.2 & 150.8 \\ 
        (1.25,0.25) & ~ & 4.3 & 4.7 & 4.6 & 4.5 & 3.4 & ~ & 16.3 & 12.4 & 11.3 & 10.6 & 9.3 & ~ & 27.3 & 20.4 & 18.7 & 17.3 & 14.6 & ~ & 42.4 & 31.4 & 28.6 & 26.7 & 22.3 & ~ & 102.2 & 86.4 & 76.5 & 72.1 & 57.2 \\ 
        (2.00,0.25) & ~ & 5.2 & 5.9 & 5.9 & 5.7 & 4.6 & ~ & 31.1 & 24.5 & 22.3 & 20.9 & 17.8 & ~ & 48.9 & 40.2 & 37.1 & 34.2 & 29.7 & ~ & 67.0 & 60.3 & 56.5 & 52.8 & 45.8 & ~ & 64.3 & 128.9 & 129.3 & 124.8 & 111.2 \\ 
        (0.25,0.75) & ~ & 30.9 & 23.6 & 19.8 & 19.0 & 11.2 & ~ & 105.7 & 50.2 & 39.7 & 33.6 & 23.6 & ~ & 126.8 & 58.8 & 49.5 & 41.4 & 28.8 & ~ & 146.4 & 72.7 & 56.0 & 48.6 & 35.4 & ~ & 171.1 & 104.7 & 84.8 & 70.3 & 49.1 \\ 
        (0.75,0.75) & ~ & 53.3 & 49.1 & 43.3 & 36.7 & 17.0 & ~ & 150.9 & 80.0 & 64.8 & 54.0 & 37.7 & ~ & 165.0 & 94.4 & 76.6 & 66.3 & 45.8 & ~ & 175.4 & 112.2 & 94.3 & 77.3 & 54.9 & ~ & 165.0 & 156.9 & 130.1 & 113.9 & 79.3 \\ 
        (1.25,0.75) & ~ & 64.1 & 97.3 & 86.8 & 76.7 & 27.3 & ~ & 169.1 & 122.8 & 99.7 & 89.6 & 62.3 & ~ & 151.9 & 140.0 & 119.1 & 104.6 & 74.4 & ~ & 146.3 & 164.7 & 136.3 & 121.9 & 90.7 & ~ & 96.9 & 183.3 & 178.2 & 164.8 & 124.8 \\ 
        (2.00,0.75) & ~ & 36.5 & 127.9 & 144.4 & 147.9 & 54.0 & ~ & 84.4 & 169.8 & 167.4 & 151.2 & 119.9 & ~ & 62.4 & 162.8 & 174.9 & 171.6 & 141.7 & ~ & 49.6 & 163.0 & 173.3 & 178.5 & 156.6 & ~ & 34.4 & 114.8 & 153.8 & 171.0 & 181.6 \\ 
        (0.25,1.25) & ~ & 44.9 & 129.3 & 166.6 & 188.9 & 148.4 & ~ & 60.8 & 140.3 & 162.2 & 178.1 & 176.3 & ~ & 71.3 & 152.7 & 176.3 & 180.9 & 163.3 & ~ & 79.7 & 165.0 & 181.8 & 179.4 & 160.3 & ~ & 104.1 & 186.6 & 183.0 & 171.4 & 121.8 \\ 
        (0.75,1.25) & ~ & 31.4 & 106.8 & 140.7 & 169.4 & 122.2 & ~ & 45.0 & 104.0 & 130.4 & 148.9 & 176.1 & ~ & 50.9 & 117.9 & 143.0 & 161.4 & 177.6 & ~ & 58.7 & 137.2 & 158.1 & 173.0 & 176.4 & ~ & 72.9 & 160.5 & 177.2 & 180.4 & 167.9 \\ 
        (1.25,1.25) & ~ & 21.9 & 75.5 & 98.9 & 122.7 & 93.0 & ~ & 30.8 & 74.7 & 98.7 & 113.8 & 152.5 & ~ & 35.6 & 87.9 & 113.6 & 128.9 & 169.9 & ~ & 40.3 & 103.6 & 131.0 & 142.5 & 177.3 & ~ & 52.2 & 129.3 & 148.3 & 165.4 & 189.1 \\ 
        (2.00,1.25) & ~ & 11.9 & 44.9 & 59.3 & 76.3 & 58.3 & ~ & 18.1 & 46.8 & 61.5 & 70.9 & 104.6 & ~ & 20.2 & 53.9 & 67.5 & 84.5 & 119.0 & ~ & 23.3 & 61.3 & 78.5 & 92.7 & 131.5 & ~ & 27.2 & 74.4 & 99.1 & 113.8 & 152.5 \\ 
        (0.25,2.00) & ~ & 11.0 & 28.2 & 36.9 & 45.2 & 25.2 & ~ & 18.6 & 39.2 & 48.1 & 53.4 & 74.0 & ~ & 24.3 & 52.5 & 62.6 & 71.8 & 94.6 & ~ & 31.0 & 65.9 & 78.4 & 88.6 & 117.1 & ~ & 53.3 & 113.2 & 128.9 & 142.5 & 157.8 \\ 
        (0.75,2.00) & ~ & 8.8 & 23.2 & 29.5 & 37.0 & 21.6 & ~ & 15.8 & 33.0 & 40.2 & 47.1 & 65.0 & ~ & 20.9 & 43.6 & 53.0 & 60.0 & 82.5 & ~ & 25.8 & 54.3 & 66.0 & 77.1 & 103.3 & ~ & 44.0 & 94.2 & 113.3 & 126.4 & 154.0 \\ 
        (1.25,2.00) & ~ & 7.4 & 18.7 & 24.5 & 13.8 & 18.6 & ~ & 13.0 & 28.1 & 33.7 & 39.7 & 53.0 & ~ & 16.6 & 36.4 & 44.3 & 51.6 & 68.3 & ~ & 21.3 & 46.2 & 57.3 & 64.5 & 85.7 & ~ & 34.4 & 77.7 & 93.2 & 108.5 & 141.5 \\ 
        (2.00,2.00) & ~ & 5.8 & 15.4 & 19.0 & 10.8 & 14.6 & ~ & 9.8 & 21.0 & 25.7 & 30.7 & 42.3 & ~ & 12.3 & 27.5 & 32.8 & 39.3 & 54.1 & ~ & 14.8 & 34.3 & 42.3 & 50.0 & 66.6 & ~ & 22.1 & 54.7 & 68.6 & 80.3 & 109.3 \\ 
         \hline
        (0.25,0.25) & ~ & 2.4 & 2.3 & 2.2 & 2.2 & 2.1 & ~ & 3.9 & 3.0 & 2.8 & 2.6 & 2.3 & ~ & 5.7 & 3.9 & 3.5 & 3.2 & 2.6 & ~ & 9.0 & 5.5 & 4.8 & 4.3 & 3.2 & ~ & 30.9 & 16.0 & 12.9 & 10.5 & 6.3 \\ 
        (0.75,0.25) & ~ & 2.7 & 2.4 & 2.4 & 2.3 & 2.3 & ~ & 5.6 & 4.1 & 3.7 & 3.5 & 3.0 & ~ & 9.0 & 6.1 & 5.5 & 4.8 & 3.9 & ~ & 15.1 & 9.5 & 8.3 & 7.3 & 5.5 & ~ & 49.4 & 32.0 & 26.1 & 22.0 & 14.3 \\ 
        (1.00,1.00) & ~ & 7.1 & 157.6 & 198.4 & 177.2 & 92.4 & ~ & 13.4 & 161.9 & 198.5 & 166.0 & 87.5 & ~ & 17.4 & 155.9 & 196.6 & 171.7 & 90.9 & ~ & 19.8 & 168.2 & 198.9 & 177.6 & 88.4 & ~ & 19.4 & 166.2 & 197.3 & 173.0 & 90.7 \\ 
        (1.25,0.25) & ~ & 3.0 & 2.8 & 2.7 & 2.7 & 2.5 & ~ & 7.6 & 6.2 & 5.7 & 5.1 & 4.3 & ~ & 13.7 & 10.4 & 9.2 & 8.2 & 6.4 & ~ & 22.3 & 17.6 & 14.9 & 13.4 & 10.0 & ~ & 38.5 & 58.5 & 50.7 & 43.4 & 30.9 \\ 
        (2.00,0.25) & ~ & 2.4 & 3.6 & 3.6 & 3.4 & 3.2 & ~ & 2.7 & 13.0 & 12.0 & 11.1 & 9.1 & ~ & 2.9 & 22.9 & 21.4 & 19.8 & 16.1 & ~ & 2.0 & 36.0 & 35.6 & 32.6 & 26.5 & ~ & 1.4 & 71.1 & 85.1 & 92.6 & 80.1 \\ 
        (0.25,0.75) & ~ & 13.6 & 5.4 & 4.4 & 3.8 & 2.9 & ~ & 56.9 & 14.8 & 10.3 & 7.6 & 4.4 & ~ & 80.2 & 20.3 & 14.2 & 10.2 & 5.4 & ~ & 103.4 & 27.4 & 18.8 & 13.5 & 6.7 & ~ & 153.9 & 53.3 & 34.5 & 24.2 & 10.8 \\ 
        (0.75,0.75) & ~ & 25.4 & 10.1 & 7.5 & 6.2 & 4.2 & ~ & 98.4 & 34.7 & 23.4 & 17.7 & 9.3 & ~ & 131.8 & 48.0 & 33.1 & 24.2 & 12.8 & ~ & 148.0 & 63.0 & 43.7 & 33.1 & 16.9 & ~ & 99.8 & 113.3 & 82.6 & 62.5 & 30.4 \\ 
        (1.25,0.75) & ~ & 36.6 & 22.1 & 15.7 & 12.2 & 7.4 & ~ & 69.2 & 75.7 & 55.3 & 40.8 & 22.6 & ~ & 65.4 & 100.0 & 74.9 & 55.8 & 30.8 & ~ & 50.5 & 126.4 & 95.6 & 76.1 & 41.0 & ~ & 14.6 & 167.3 & 148.9 & 128.2 & 74.5 \\ 
        (2.00,0.75) & ~ & 17.3 & 67.6 & 51.6 & 39.5 & 22.5 & ~ & 1.3 & 130.4 & 135.3 & 119.2 & 76.2 & ~ & 1.3 & 112.5 & 147.4 & 144.1 & 101.3 & ~ & 1.3 & 91.3 & 145.2 & 147.7 & 124.0 & ~ & 1.4 & 36.2 & 83.4 & 131.3 & 172.2 \\ 
        (0.25,1.25) & ~ & 3.8 & 38.6 & 61.3 & 78.4 & 132.7 & ~ & 16.5 & 94.8 & 129.4 & 149.6 & 142.9 & ~ & 22.8 & 115.1 & 149.9 & 165.1 & 114.7 & ~ & 29.4 & 132.9 & 163.3 & 169.2 & 89.8 & ~ & 47.2 & 171.6 & 163.9 & 122.7 & 38.7 \\ 
        (0.75,1.25) & ~ & 2.3 & 20.3 & 33.5 & 46.3 & 86.3 & ~ & 6.5 & 54.7 & 81.8 & 107.6 & 163.5 & ~ & 10.7 & 68.8 & 100.6 & 128.4 & 160.1 & ~ & 14.5 & 85.2 & 121.2 & 148.4 & 163.5 & ~ & 22.1 & 125.4 & 161.1 & 178.2 & 123.1 \\ 
        (1.25,1.25) & ~ & 1.6 & 10.3 & 17.5 & 25.8 & 52.0 & ~ & 2.4 & 29.2 & 46.2 & 65.8 & 113.4 & ~ & 3.5 & 37.4 & 57.3 & 80.2 & 134.3 & ~ & 5.3 & 46.7 & 69.0 & 96.1 & 154.5 & ~ & 8.4 & 72.0 & 108.0 & 135.2 & 177.4 \\ 
        (2.00,1.25) & ~ & 1.3 & 3.9 & 6.6 & 10.2 & 21.8 & ~ & 1.2 & 9.1 & 16.2 & 24.6 & 52.7 & ~ & 1.2 & 12.3 & 21.5 & 31.5 & 65.4 & ~ & 1.3 & 15.6 & 26.0 & 39.2 & 79.0 & ~ & 1.4 & 21.0 & 37.5 & 56.8 & 114.6 \\ 
        (0.25,2.00) & ~ & 1.6 & 3.7 & 4.5 & 5.4 & 7.9 & ~ & 3.1 & 12.3 & 16.9 & 21.5 & 34.7 & ~ & 5.1 & 19.2 & 26.0 & 32.9 & 51.9 & ~ & 7.8 & 27.8 & 38.0 & 47.7 & 70.8 & ~ & 19.8 & 66.5 & 83.9 & 100.9 & 107.8 \\ 
        (0.75,2.00) & ~ & 1.5 & 3.1 & 3.9 & 4.6 & 6.7 & ~ & 2.2 & 9.1 & 12.9 & 16.4 & 27.1 & ~ & 3.4 & 14.2 & 20.0 & 25.7 & 41.0 & ~ & 5.2 & 21.4 & 29.1 & 37.3 & 58.5 & ~ & 12.3 & 50.4 & 67.9 & 82.8 & 114.3 \\ 
        (1.25,2.00) & ~ & 1.3 & 2.7 & 3.4 & 4.0 & 5.6 & ~ & 1.6 & 6.6 & 9.4 & 12.4 & 21.1 & ~ & 2.1 & 10.1 & 14.6 & 18.7 & 31.5 & ~ & 2.9 & 15.5 & 21.6 & 27.2 & 45.7 & ~ & 5.9 & 34.8 & 48.4 & 62.4 & 98.3 \\ 
        (2.00,2.00) & ~ & 1.2 & 2.1 & 2.6 & 3.1 & 4.4 & ~ & 1.1 & 3.5 & 5.4 & 7.3 & 13.3 & ~ & 1.2 & 5.3 & 8.4 & 11.2 & 20.5 & ~ & 1.2 & 7.7 & 12.1 & 16.4 & 29.5 & ~ & 1.3 & 15.3 & 23.8 & 34.8 & 63.1 \\  \hline \hline 
   \end{tabular} \label{tab:log2}
\end{sidewaystable}
\newpage
\subsection{Performance Comparison}

The study conducted by \cite{Ali} introduced a control chart for monitoring the time between events based on the NHPP with power law intensity, without accounting for risk factors and the TC variable. In this subsection, we evaluate the performance of the proposed chart in comparison to Ali's chart, using the ARL metric. In the initial phase, we generate 5000 TBEs with parameters $\gamma_0=0.05$, $\eta_0=1.50$, and $\beta=0.50,2.00$ using the RANHPP model. Then the parameters are estimated using the IFM approach explained in the Appendix. For the proposed model with $\beta=0.50$, the estimated parameters are $\hat{\gamma}=0.05$, $\hat{\eta}=1.52$, and $\hat{\beta}=0.52$ that are fairly close the known values. In contrast, the NHPP-based method proposed by \cite{Ali} yields parameter estimates of $\hat{\gamma}=0.11$ and $\hat{\eta}=1.43$. These biased estimates, especially in the case of $\gamma$, are caused by the noise incurred by the risk factor that has not been accounted for in the estimation process. This unaccounted noise is expected to impact the chart's performance.\\

Given the estimated parameters, we calculated the control limits of the proposed method (by ignoring TC) from the conditional CDF of $X_i$ given in \eqref{equ:cdf_TBE} and the limits of the alternative method from  \cite{Ali}. This means that the risk factor is ignored in the design of the alternative approach. Then, the performance of both approaches is compared using the ARL metric across 17 combinations of $(\delta_\gamma,\delta_\eta)$. Table \ref{tab:compar} presents the $ARL$ values of the RANHPP-based and NHPP-based methods when monitoring TBEs in the presence of a risk factor. The crucial finding in the table is that, while Ali's approach detects some shifts sooner than the proposed method, its actual $ARL_0$ consistently falls below the target value of $200$. Specifically, the IC $ARL$ for this approach is $187.56$ and $28.65$ when $\beta=0.50$ and $\beta=2.00$, respectively.\\

Ultimately, it can be concluded that although the alternative approach outperforms in some instances, this improved performance comes at the cost of generating numerous false alarms, particularly when $\beta$ is relatively large. These frequent alarms lose their warning role and are unlikely to trigger an audit each time.\\

\begin{table}[]
    \centering
    \caption{$ARL$ values of the RANHPP and the NHPP TBE control charts.}
    \begin{tabular}{cccccc}
    \hline \hline 
         & \multicolumn{2}{c}{ $\beta=0.50$}&& \multicolumn{2}{c}{$ \beta=2.00$}\\
         \cline{2-3} \cline{5-6}
        $(\delta_{\gamma},\delta_{\eta})$ & RANHPP & NHPP & ~ & RANHPP & NHPP\\ \hline 
        (0.25,0.25) & 1.06 & 1.07 & ~ & 1.33 & 1.54 \\ 
        (0.75,0.25) & 1.19 & 1.20 & ~ & 1.65 & 2.08 \\ 
        (1.00,1.00) & 198.98 & 187.56 & ~ & 197.43 & 28.65 \\ 
        (1.25,0.25) & 1.28 & 1.32 & ~ & 1.92 & 2.38 \\ 
        (2.00,0.25) & 1.41 & 1.46 & ~ & 2.17 & 2.83 \\ 
        (0.25,0.75) & 1.62 & 1.76 & ~ & 2.21 & 4.28 \\ 
        (0.75,0.75) & 3.60 & 4.20 & ~ & 6.13 & 30.73 \\ 
        (1.25,0.75) & 6.64 & 7.95 & ~ & 13.24 & 45.59 \\ 
        (2.00,0.75) & 13.37 & 16.72 & ~ & 30.21 & 35.01 \\ 
        (0.25,1.25) & 196.56 & 146.54 & ~ & 142.16 & 27.37 \\ 
        (0.75,1.25) & 105.7 & 87.95 & ~ & 141.2 & 19.42 \\ 
        (1.25,1.25) & 73.28 & 64.16 & ~ & 102.2 & 17.18 \\ 
        (2.00,1.25) & 55.68 & 44.79 & ~ & 72.51 & 14.75 \\ 
        (0.25,2.00) & 29.12 & 26.18 & ~ & 48.05 & 13.52 \\ 
        (0.75,2.00) & 20.58 & 18.4 & ~ & 35.76 & 11.92 \\ 
        (1.25,2.00) & 17.61 & 16.32 & ~ & 30.72 & 10.78 \\ 
        (2.00,2.00) & 15.35 & 41.32 & ~ & 26.54 & 9.16 \\ \hline \hline 
    \end{tabular} \label{tab:compar}
\end{table}

\subsection{The impact of ignoring the risk factor}
We further investigate the impact of the risk factors on the IC performance of the proposed chart that has been designed with ignorance of the risk factor. In other words, we are interested in understanding the consequences of a risk factor affecting TBE that has not been considered in the design of the corresponding chart. To do this, we follow the \textbf{Algorithm 1} except for steps 10 and 11 where we calculate the control limits from the CDF of $W_i$ in \eqref{equ:Wdist} that is not conditioned on $\bm{z}_{i}$. The simulation setting is based on the parameters mentioned earlier in this section with a nominal $ARL_0=200$.  The results of this study (not presented here) showed that the actual false alarm rate of the chart is considerably higher than its nominal value. This observation aligns with our findings when comparing the results with those of \cite{Ali}. For example, the $ARL_0$ values corresponding to $\beta=-2.00, -0.50, 0.50, 2.00$ are $6.83, 111.72, 96.37,$  and $4.33$, respectively. These findings highlight the noticeable influence of neglecting the risk factors on the statistical properties of the monitoring technique.

\section{A case study: Monitoring pipeline accidents}
 \label{sec:Example}
  In this section, our objective is to apply the suggested control chart within an SPM context for the analysis of pipeline accident data briefly introduced in Section \ref{sec:ME}. The dataset in question contains an extensive set of oil pipeline accidents, including leaks and spills. Our primary goal is to explore how the proposed method can be effectively utilized and demonstrate the steps involved in its implementation using this dataset as a case study.

    \subsection{Data description}
    Following the details outlined in Section \ref{sec:ME}, it encompasses a collection of 2789 records documenting oil pipeline accidents, encompassing leaks or spills, which were reported to the PHMSA during the period spanning from 2010 to 2017. While the letter-value plots in Figure \ref{fig:TBE-cost} (a) and (b) show a few outliers in TBEs and TCs, we decided to keep them and apply the proposed method. Table \ref{tab:100fails} presents the information on the last 50  accidents.\\
    
  The accident date variable denotes the precise date of the oil spill, and it will serve as the basis for calculating the TBE, represented as $X_i$ and measured in hours. The TC variable represents the sum of various costs, including Property Damage Costs, Lost Commodity Costs, Public/Private Property Damage Costs, Emergency Response Costs, Environmental Remediation Costs, and Other Costs  (which be measured by 100,000 dollars), and will be denoted by $Y_i$. Additionally, there are five confounding variables that will be used as risk factors which are shown in Table \ref{tab:covar}. \\
  
\begin{figure}
\centering
\subfigure[TBE $X_i.$]{%
\resizebox*{15cm}{!}{\includegraphics[width=15cm,height=4.5cm]{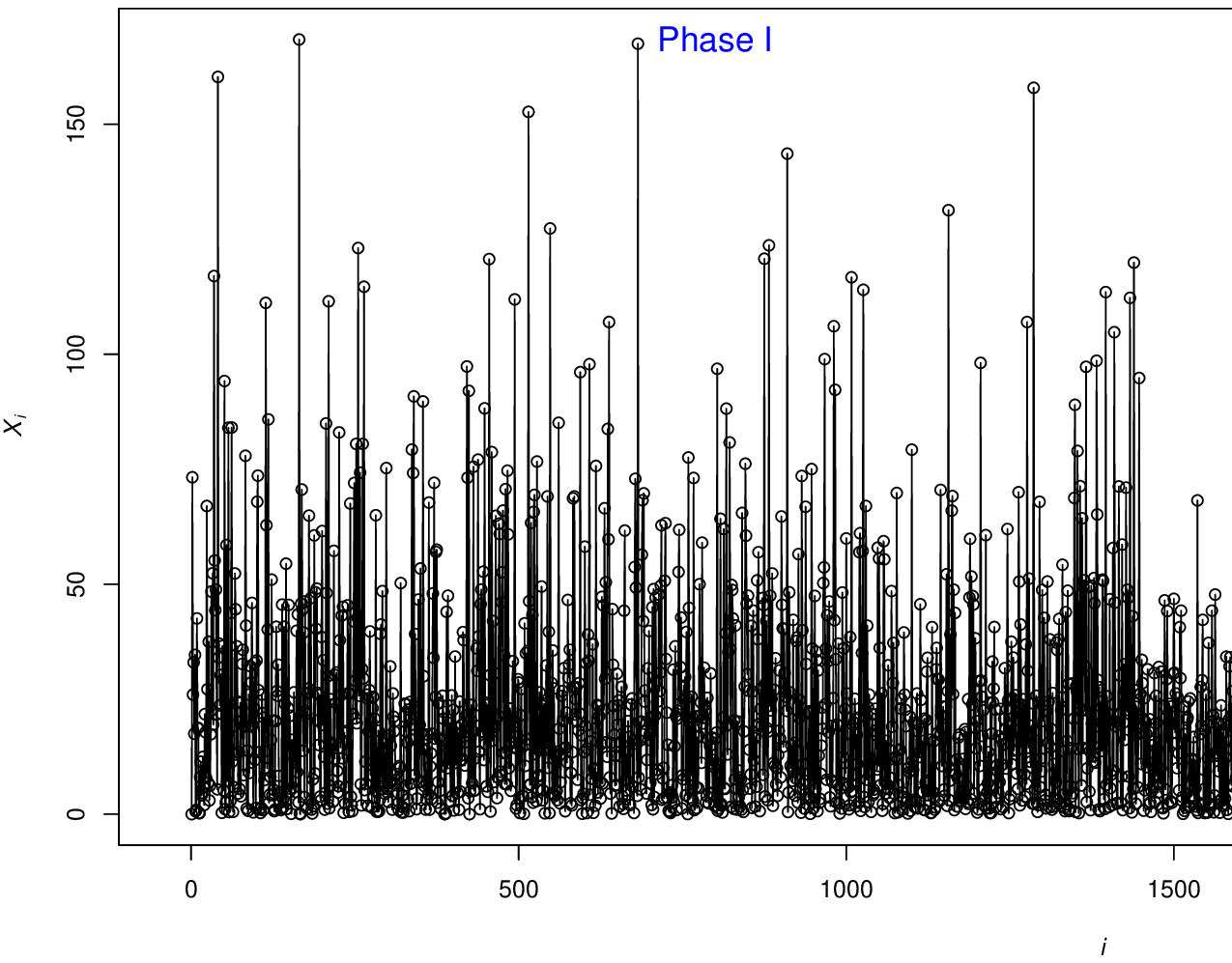}}}
\subfigure[TC $Y_i$.]{%
\resizebox*{15cm}{!}{\includegraphics[width=15cm,height=4.5cm]{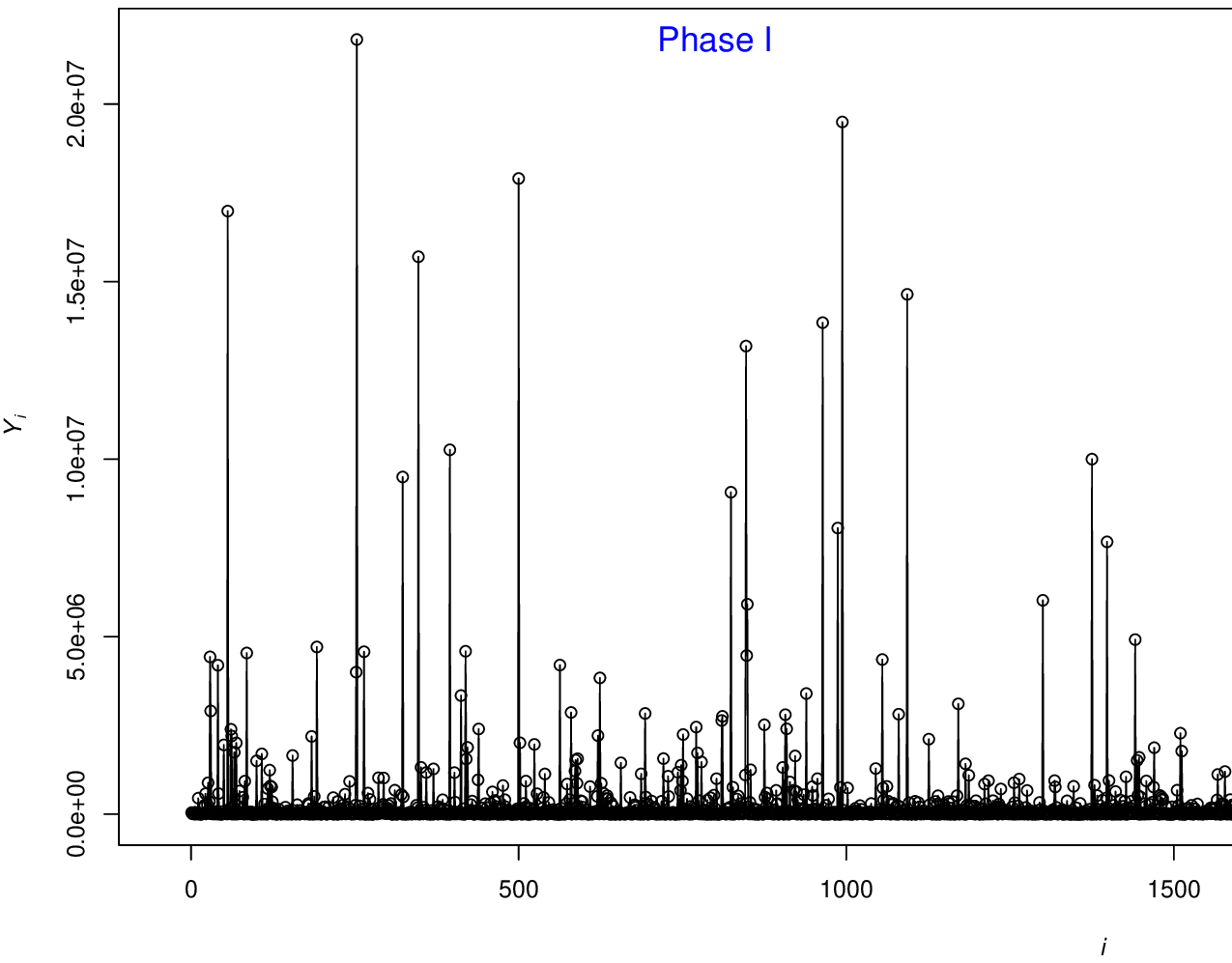}}}
\subfigure[AC $W_i$.]{%
\resizebox*{15cm}{!}{\includegraphics[width=15cm,height=4.5cm]{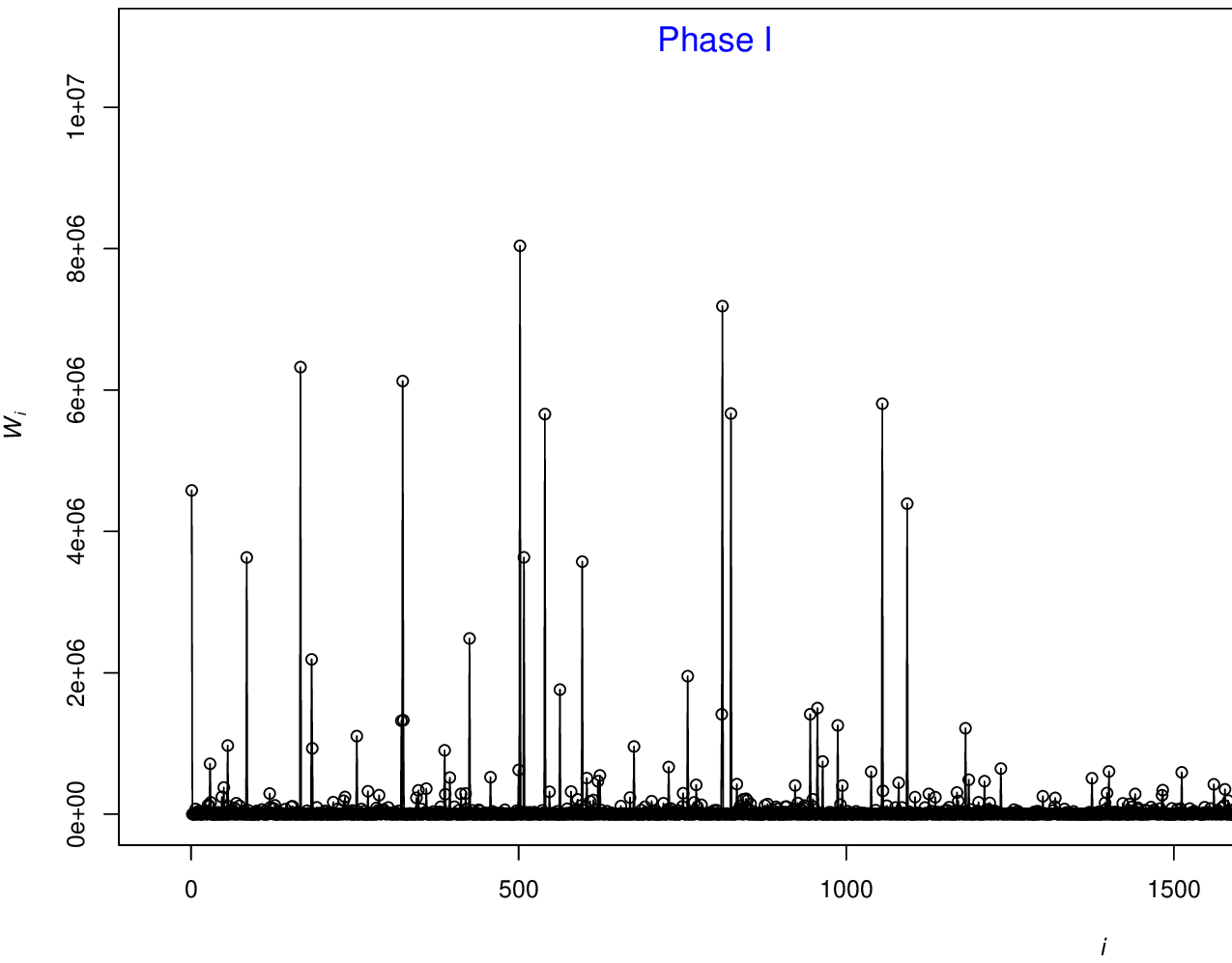}}}
\caption{Pipelines accidents raw data.} \label{fig:plot}
\end{figure}

Panels (a)-(c) of Figure \ref{fig:plot} show the observations of the $X$ (TBE)  and $Y$ (TC) variables along with the $W$'s against the time index $i$.  In this Figure, panel (a) reveals that a substantial portion of the TBE falls within the 0 to 50-hour range. Meanwhile, panel (b) illustrates that the total cost of failures typically ranges from 0 to 5 million USD. Consequently, this suggests that the AC for most failures is close to $100,000$ USD, as depicted in panel (c) of Figure \ref{fig:plot}. \\

   \begin{table}[]
    \caption{Information on risk factors.}\label{tab:covar}
    \centering
    \begin{tabular}{p{5cm}p{9cm}}
        \hline\hline
        Covariate & Description \\
        \hline
        Pipeline Location (PL) & Classifies pipelines into two categories: (1) onshore (ONS) or (2) offshore (OFS). \\
        
        Pipeline Type (PT) & Categorizes pipelines as follows: (1) Aboveground (ABG), (2) Underground (UNG), (3) Tank (TAN), or (4) Transition Area pipelines (TRA). \\
        
        Liquid Type (LT) & Discriminates between various liquid oil types, including (1) Biofuel/Alternative Fuel (BAF), (2) CO2, (3) Crude Oil (CRO), (4) HVL or Other Flammable or Toxic Fluid, GA (HFT), and (5) Refined and/or Petroleum Product (RPL). \\
        
                Accident State (AS) & The report provides data for forty-six states across the US. \\
                
        Oil Cause Categories (CC) & Encompasses the following categories: (1) All Other Causes (AOC), (2) Other Outside Force Damage (OFD), (3) Corrosion (COR), (4) Excavation Damage (EXD), (5) Incorrect Operation (INO), (6) Material/Weld/Equip Failure (MWE), and (7) Natural Force Damage (NFD). \\
       
 \hline\hline
    \end{tabular}
\end{table}

\subsection{Estimating the process parameters}

For illustration purposes, we randomly selected the last 100 data points for monitoring purposes in Phase II and the rest of them are used to estimate the process parameters. To estimate the parameters, we employed the IFM approach explained in the Appendix. In the initial step, we employed the HPP as well as both power-law and log-linear models to fit the TBE variable based on a RANHPP and chose the model based on the Akaike Information Criterion (AIC). Among them, the power-law model yielded a better fit. The estimated parameters are in Table \ref{tab:phaseI}  where $\hat{\beta}_1, \hat{\beta}_2, \hat{\beta}_3, \hat{\beta}_4$, and $\hat{\beta}_5$ are the regression coefficients of PL, PT, LT, AS and CC, respectively. The hazard ratios (HR) defined as $e^\beta$ regarding risk factors are also reported in this table. In this context, HR quantifies the multiplicative change in the hazard rate for a one-unit increase in the risk factor. In this way, HR$>1 \quad (<1)$ indicates an increase (decrease) in the hazard rate associated with a one-unit change in the risk factor. Furthermore, $HR=1$ implies that there is no change in the hazard rate associated with a one-unit change in the predictor variable. \\

Based on the table, $\hat{\eta}>1$ indicates that the hazard rate of the pipelines is increasing over time, which aligns with our expectation given their degradation nature. This translates to a higher frequency of failures or a shorter TBE in the presence of aging. Furthermore, the estimated coefficients of the risk model show that there is a negative association between the risk factors PL and the TBE variable  ($\beta_1<0$ with corresponding HR $<1$), whereas covariates PT, LT, AS, and CC are positively associated with TBE ($\beta_2>0, \beta_3>0, \beta_4>0$, and $\beta_5>0$ with corresponding HR $>1$). Furthermore, according to the HR criteria, we can see that the risk factor PL (pipeline location) has the most effect on TBE compared to the other risk factors.\\

 Now, turning to the TC variable denoted by $Y$. To assess the distribution of $Y_i$'s, a Kolmogorov-Smirnov test with corresponding $p$-value$=0.025$ suggests that an exponential distribution with mean $\frac{1}{\hat{\mu}}=311,905$ USD could be a reasonable model for the TC variable.  In the second step of our analysis, the focus shifts to selecting the most suitable copula based on the estimated parameters obtained in the first step. We search through three Archimedean copula functions including, Gumbel, Frank and Clayton to find the most suitable model for describing the dependence structure between variables $X_i|T_{i-1},\bm{z_{i}}$ and $Y_i$.  Based on the AIC metric, the Gumbel copula with $\tau=0.001$ is the optimal choice that indicates a weak positive dependence between the TBE and TC variables. Eventually, the estimated parameters in Table \ref{tab:phaseI} can be used to establish the control limits of the proposed approach in Phase II to monitor the AC over time through the ratios $W_i=\frac{Y_i}{X_i}$ for $i=1,2,\ldots$.\\

\begin{table}
    \centering
    \footnotesize
    \caption{Estimated parameters of the power law RANHPP model for pipeline accidents data.}\label{tab:phaseI}
    \begin{tabular}{cccccccccccc}
     \hline\hline
    &  $\hat{\gamma}$ & $\hat{\eta}$ & $\hat{\beta}_1$ & $\hat{\beta}_2$ & $\hat{\beta}_3$ & $\hat{\beta}_4$ & $\hat{\beta}_5$ & $\hat{\mu}$& Copula&$\hat{\tau}$&AIC \\  \hline
%
       & 0.024&  1.071& -0.131& 0.002& 0.016& 0.001&  0.010 &  3.21e-06& Gumbel & 0.001& -95039.97  \\ 
       HR $\rightarrow$ &- & - & 0.877& 1.002&1.016 & 1.001 & 1.010&- & - & &  \\ 
       \hline\hline
        \end{tabular}
\end{table}

\subsection{Phase II analysis}
 
\begin{figure}[H]
\centering
\subfigure[Control chart based on original observations.]{%
\resizebox*{15cm}{!}{\includegraphics[width=15cm,height=4.45cm]{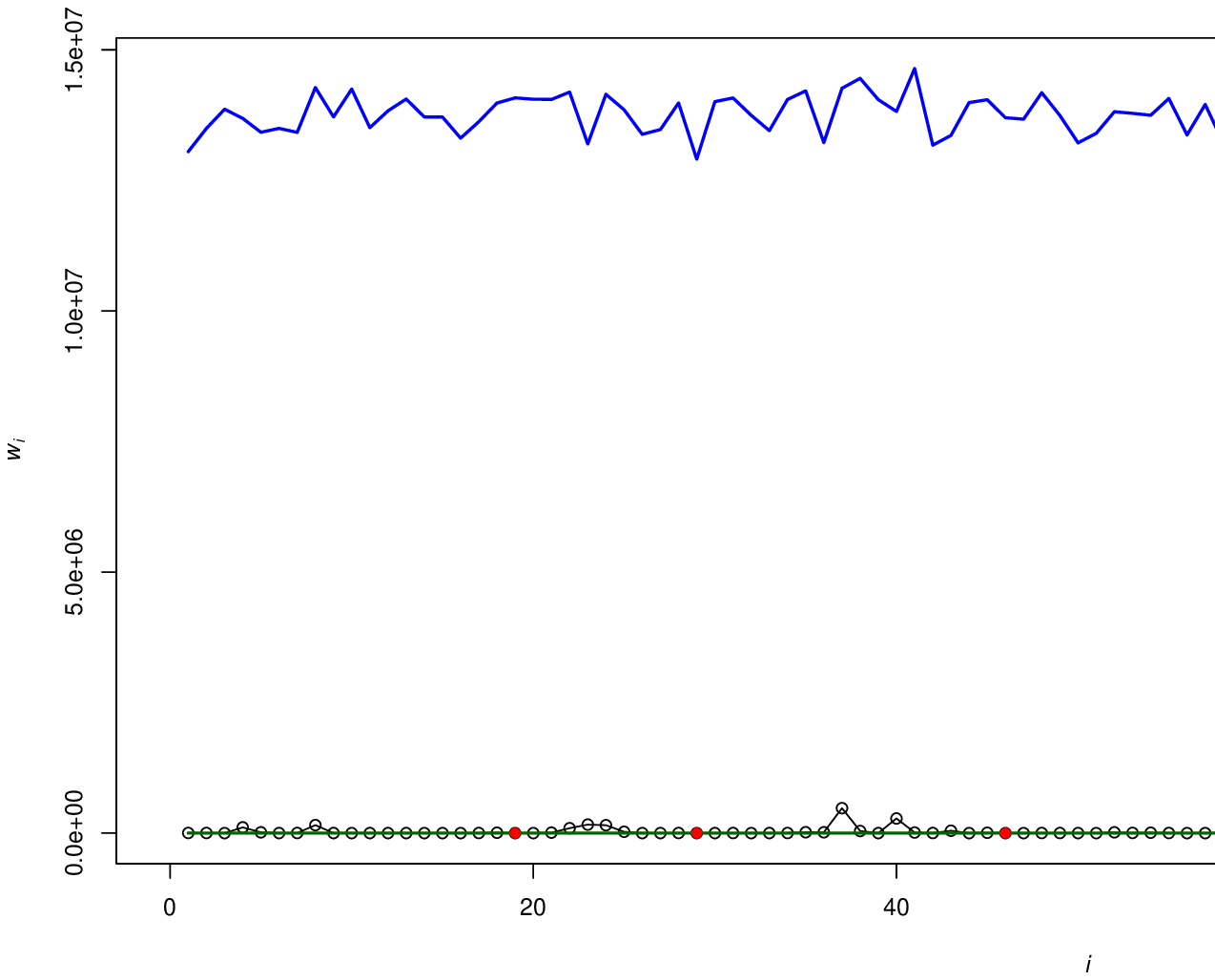}}}
\subfigure[Control chart for $W_i$ based on log-transformed observations and control limits.]{%
\resizebox*{15cm}{!}{\includegraphics[width=15cm,height=4.45cm]{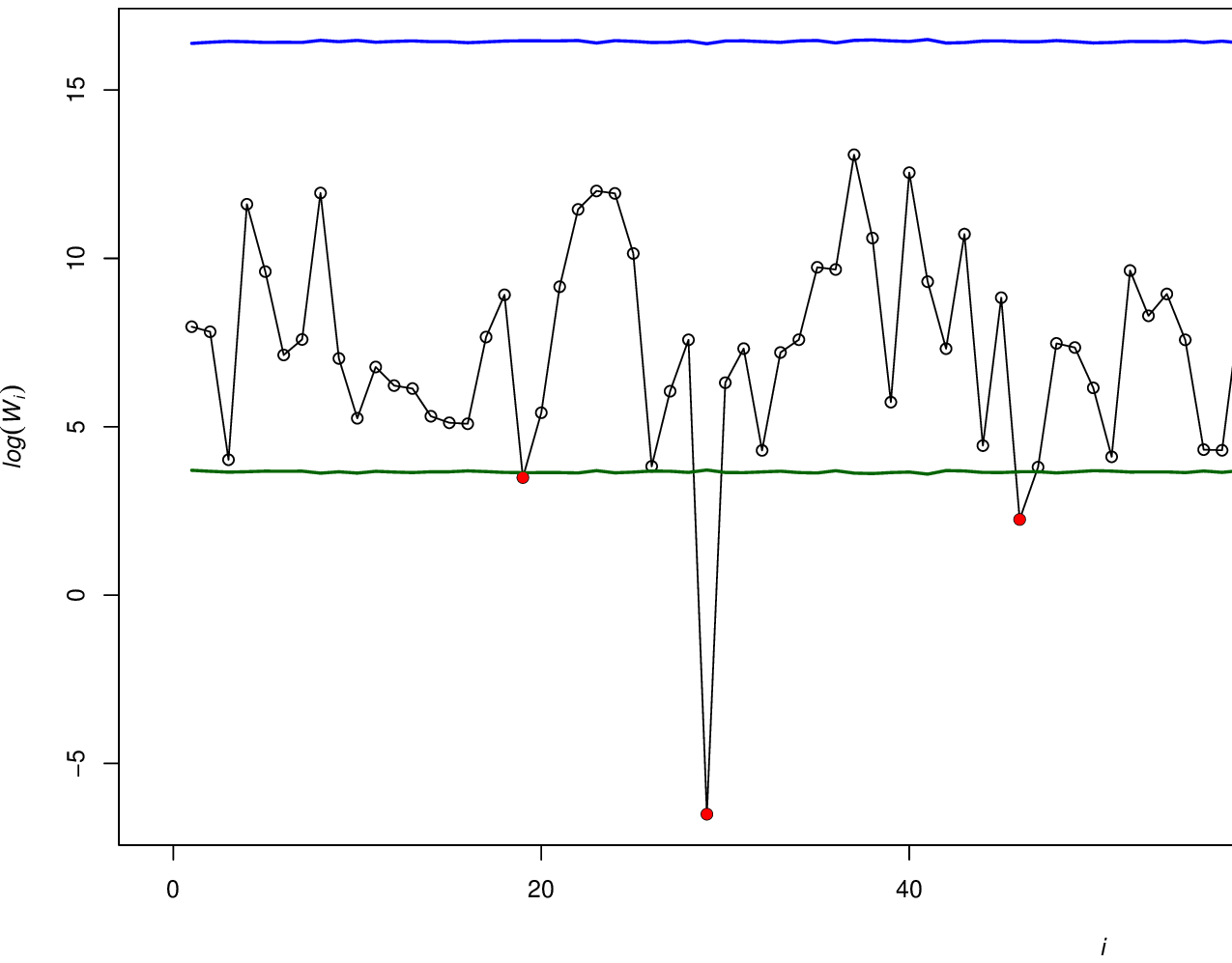}}}
\subfigure[Control chart for $Y_i$ based on log-transformed observations and control limits.]{%
\resizebox*{15cm}{!}{\includegraphics[width=15cm,height=4.45cm]{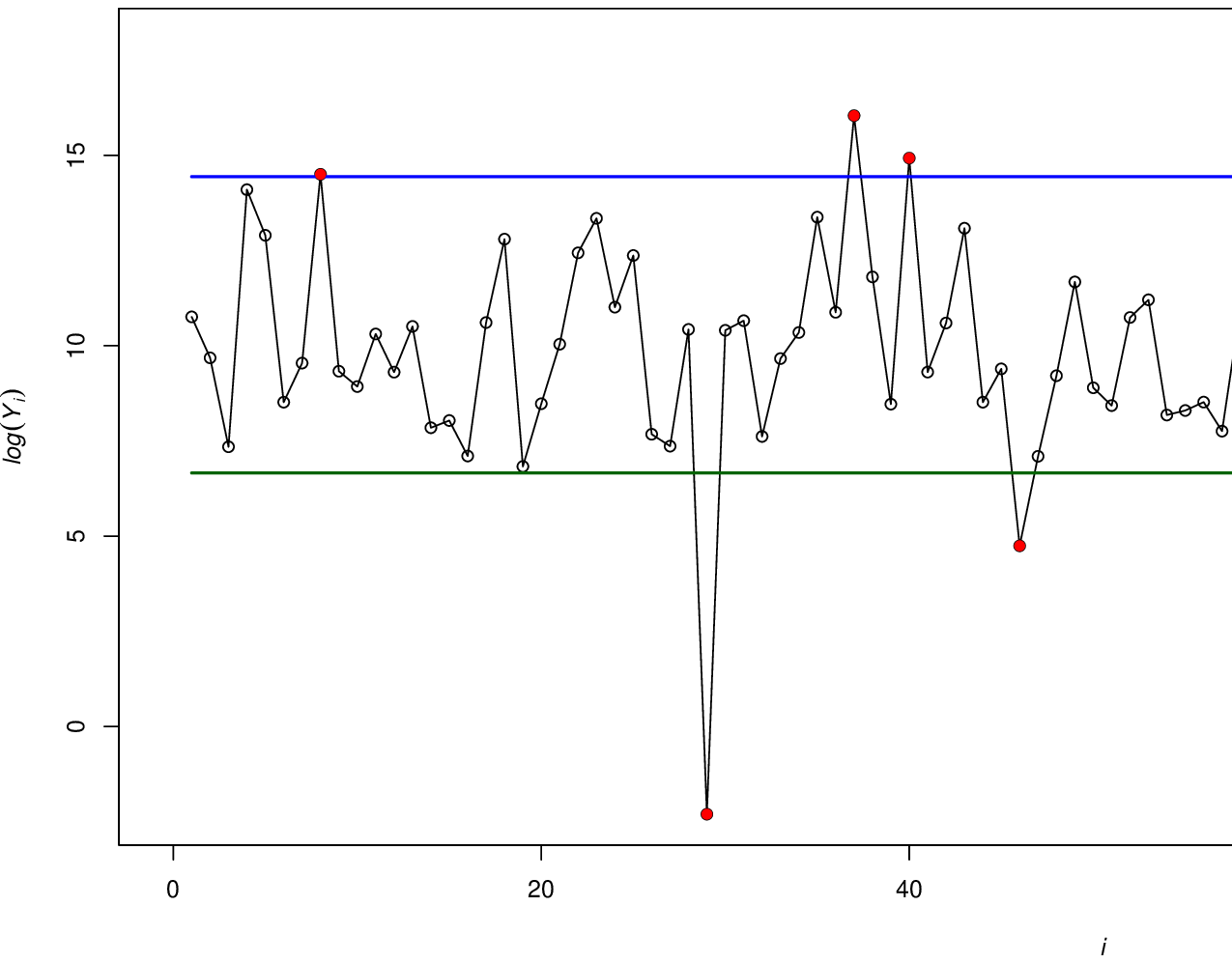}}}
\subfigure[Control chart for $X_i$ based on log-transformed observations and control limits.]{%
\resizebox*{15cm}{!}{\includegraphics[width=15cm,height=4.5cm]{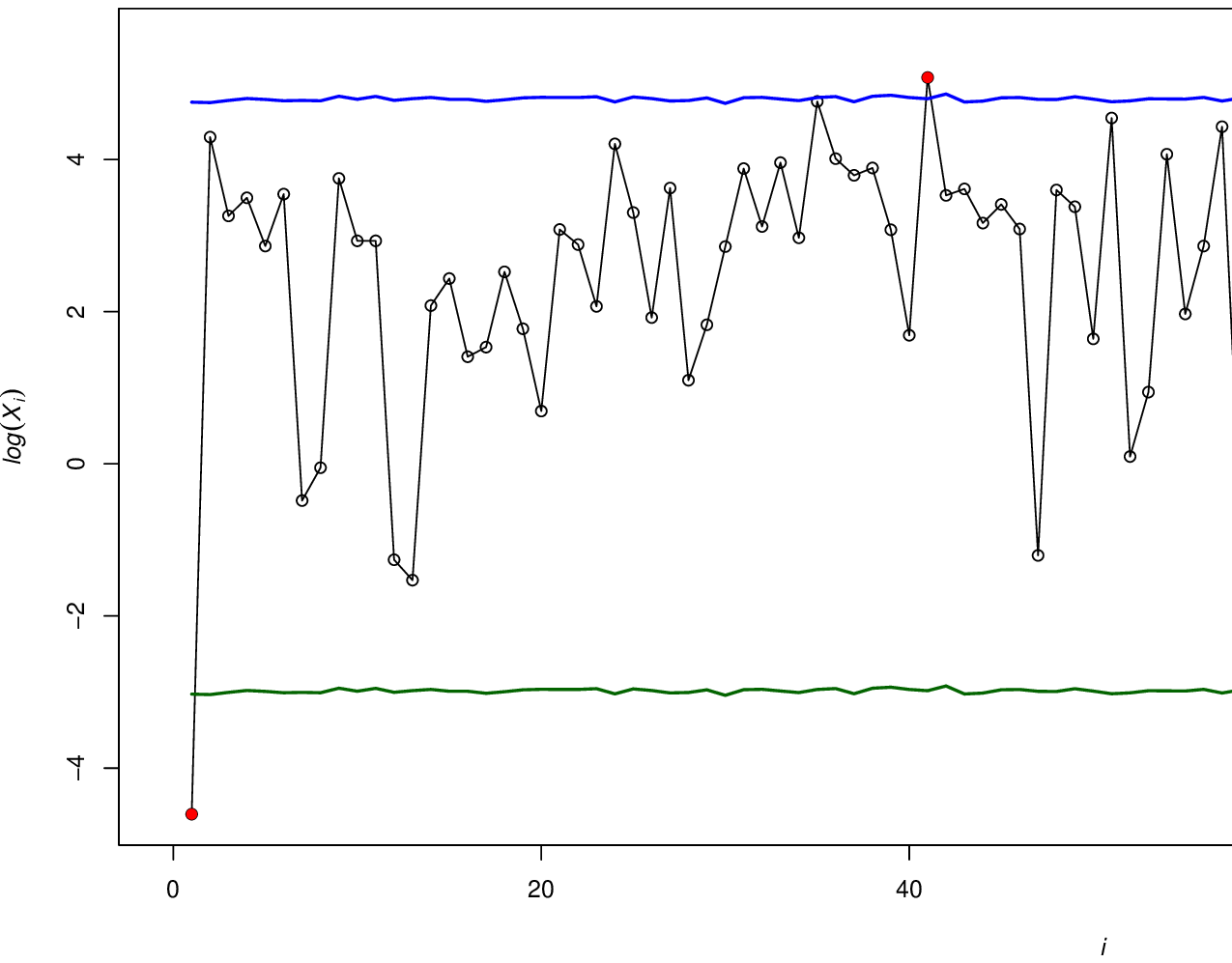}}}
\caption{Control chart for pipeline accidents in Phase II.} \label{fig:Phase II}
\end{figure}

To calculate the control limits in this phase, we set $ARL_0=200$ corresponding to the false alarm rate $\alpha=0.005$. Figure \ref{fig:Phase II}(a) depicts the control chart representing the last one hundred observations of $W_i$. Since the distribution of $W_i$ is highly right-skewed, is hard to conclude the stability of the process from this figure. Thus, to better illustrate, Figure \ref{fig:Phase II}(b) shows the control chart after taking a logarithm of the monitor statistics and their corresponding control limits. The updated figure makes it much easier to assess the random behaviour of $W_i$ within the IC zoon and to highlight its OC observations. Figure \ref{fig:Phase II}(b) shows nine OC signals by plotting $W_i$ below the $LCL_i$.  For more illustration, Table \ref{tab:100fails} presents the information of the last 50 failures. The OC signals are shown by bold-red values in this table. Considering all this information, the proposed control chart suggests evidence of a potential decrease (improvement) in the AC parameter.  \\

Although the control chart for AC provides valuable insights into the process, it does not address whether the TBE, the TC, or both of them have shifted. This is a known disadvantage of control charts based on more than characteristics. In such cases, one idea is to develop control charts for monitoring TBE and TC individually to find the main source of shift in the AC. Figure \ref{fig:Phase II}(c) and (d) show separate control charts for $X_i$ and $Y_i$. Note that the control limits of TBE and TC charts are obtained based on the quantiles of the conditional CDF of $X_i$ in \eqref{equ:cdf_TBE} and the quantiles of the CDF of the exponential distribution, respectively. Figure \ref{fig:Phase II}(d) indicates that the TBE variable remains relatively stable statistically throughout this period. However, in panel (c), the control chart of TC displays several OC signals, particularly towards the end of the monitoring period. This coincides with the presence of OC signals on the AC chart (panel (a) of this figure). On the other hand, the TC's control chart shows a downward shift in $Y_i$, which is the same as the direction of AC's shift. By analyzing the insights provided by both the joint and individual control charts, it can be concluded that the downward deviation in TC is the root cause of the observed OC condition in the AC's control chart. \\

It is important to note that while a downward shift in AC/TC is a positive sign, the inspectors should continue the monitoring process for some time to ensure that the improvement is sustained and that other aspects of the process are not adversely affected. Since the TC is the aggregate sum of all incurred costs, the manager can investigate what specific cost factor(s) have contributed to the decrease in TC. In case a sustained decrease in TC is approved, the quality inspector should plan to collect data and reestablish control limits based on new data.\\

\begin{table}[!ht]
    \centering
    \setlength{\tabcolsep}{5pt}  
\renewcommand{\arraystretch}{1}  
    \small
    \caption{The information on the last fifty accidents.}
    \begin{tabular}{cccccccccccc}
    \hline\hline
        i & $T_i$ & $X_i$ & $Y_i$ & $W_i$ & $LCL_i$ & $UCL_i$ & $Z_{1i}$ & $Z_{2i}$ & $Z_{3i} $& $Z_{4i} $& $Z_{5i}$ \\ 
        \hline
          2732 & 59562.4 & 75.2 & 4575 & 60.8 & 36.9 & 13197434.0 & ONS & ABG & CRO & OH & MWE \\ 
        2733 & 59565.4 & 3.0 & 46280 & 15341.4 & 37.0 & 13168685.0 & ONS & UNG & HFT & IL & MWE \\ 
        2734 & 59583.8 & 18.4 & 73569 & 3998.3 & 37.1 & 13132831.0 & ONS & ABG & RPL & CA & MWE \\ 
        2735 & 59584.2 & 0.5 & 3571 & 7652.1 & 36.3 & 13439985.0 & ONS & ABG & CRO & KS & MWE \\ 
        2736 & 59586.3 & 2.1 & 4020 & 1961.0 & 38.1 & 12772968.0 & ONS & UNG & HFT & TX & MWE \\ 
        2737 & 59652.9 & 66.7 & 5011 & 75.2 & 36.6 & 13330405.0 & ONS & TAN & RPL & NJ & AOC \\ 
        2738 & 59684.6 & 31.7 & 2336 & 73.7 & 38.5 & 12624830.0 & ONS & TAN & RPL & SD & MWE \\ 
        2739 & 59695.9 & 11.3 & 60000 & 5301.9 & 37.3 & 13084386.0 & ONS & ABG & CRO & TX & MWE \\ 
        2740 & 59712.1 & 16.2 & 157221 & 9725.0 & 36.7 & 13303158.0 & ONS & ABG & CO2 & TX & MWE \\ 
        2741 & 59769.9 & 57.8 & 4120 & 71.3 & 36.4 & 13410679.0 & ONS & UNG & CRO & TX & COR  \\ 
        2742 & 59871.1 & 101.2 & 4521 & 44.7 & 37.3 & 13081148.0 & ONS & ABG & CRO & TX & MWE \\ 
        2743 & 59892.1 & 21.1 & 5417 & 257.3 & 37.9 & 12853827.0 & ONS & ABG & HFT & TX & MWE \\ 
        2744 & 59894.7 & 2.6 & 145310 & 56249.0 & 35.1 & 13876192.0 & ONS & UNG & CRO & CA & COR \\ 
        2745 & 59903.1 & 8.4 & 1016236 & 121462.5 & 35.9 & 13564725.0 & ONS & UNG & HFT & MO & AOC \\ 
        2746 & 59914.1 & 11.1 & 3089580 & 279600.0 & 37.6 & 12966895.0 & ONS & ABG & RPL & LA & MWE \\ 
        2747 & 59936 & 21.9 & 157000 & 7174.4 & 38.5 & 12619194.0 & ONS & ABG & RPL & TX & MWE \\ 
        2748 & 59936.6 & 0.6 & 22203 & 39181.8 & 35.3 & 13810797.0 & ONS & ABG & CO2 & UT & AOC \\ 
        2749 & 59943.4 & 6.9 & 70015 & 10171.7 & 37.3 & 13079746.0 & ONS & ABG & CRO & TX & MWE \\ 
        2750 & 59962.4 & 19.0 & 825 & 43.5 & 37.1 & 13133272.0 & ONS & ABG & RPL & AR & MWE \\ 
        2751 & 59963.1 & 0.7 & 800 & 1142.9 & 38.0 & 12795613.0 & ONS & TRA & CRO & WY & NFD \\ 
        2752 & 59971.9 & 8.8 & 400 & 45.5 & 36.2 & 13460802.0 & ONS & ABG & CRO & IL & MWE \\ 
        2753 & 60039.1 & 67.2 & 7500 & 111.6 & 35.5 & 13721322.0 & ONS & UNG & CRO & ND & AOC \\ 
        2754 & 60070.2 & 31.1 & 45 & \textbf{\rd{1.4}} & 36.9 & 13216967.0 & ONS & ABG & CRO & TX & INO \\ 
        2755 & 60081.7 & 11.5 & 100500 & 8713.9 & 37.8 & 12894699.0 & ONS & ABG & HFT & SC & MWE \\ 
        2756 & 60143.9 & 62.1 & 523 & 8.4 & 37.2 & 13105401.0 & ONS & UNG & RPL & TX & AOC \\ 
        2757 & 60168.1 & 24.3 & 1820500 & 75072.2 & 38.1 & 12763518.0 & ONS & ABG & RPL & TX & INO \\ 
        2758 & 60171.1 & 3.0 & 1000 & 339.0 & 37.5 & 13006054.0 & ONS & ABG & HFT & NM & MWE \\ 
        2759 & 60173.9 & 2.8 & 27580 & 9850.0 & 38.1 & 12761980.0 & ONS & ABG & RPL & TX & INO \\ 
        2760 & 60212.8 & 38.9 & 43 & \textbf{\rd{1.1}} & 37.5 & 12990762.0 & ONS & ABG & HFT & TX & INO \\ 
        2761 & 60230.4 & 17.7 & 20200 & 1143.4 & 37.9 & 12847206.0 & ONS & ABG & HFT & TX & MWE \\ 
        2762 & 60244.8 & 14.4 & 30400 & 2111.1 & 38.0 & 12789663.0 & ONS & UNG & RPL & VA & EXD \\ 
        2763 & 60252.6 & 7.8 & 15000 & 1931.3 & 36.9 & 13214086.0 & ONS & ABG & CRO & TX & INO \\ 
        2764 & 60259.7 & 7.1 & 6400 & 903.5 & 36.0 & 13536355.0 & ONS & UNG & CRO & TX & AOC \\ 
        2765 & 60281.8 & 22.1 & 119829 & 5418.0 & 38.6 & 12554030.0 & ONS & TAN & RPL & OH & NFD \\ 
        2766 & 60322.6 & 40.8 & 9450 & 231.6 & 36.9 & 13211372.0 & ONS & TAN & CRO & NJ & MWE \\ 
        2767 & 60371.4 & 48.8 & 1050 & \textbf{\rd{21.5}} & 37.3 & 13073569.0 & ONS & ABG & CRO & TX & MWE \\ 
        2768 & 60378.4 & 7.1 & 375 & 52.9 & 37.7 & 12931277.0 & ONS & ABG & CRO & TX & NFD \\ 
        2769 & 60378.8 & 0.3 & 100 & 300.0 & 35.8 & 13625136.0 & ONS & UNG & CRO & OK & AOC \\ 
        2770 & 60379.4 & 0.6 & 10993 & 18845.1 & 36.4 & 13399666.0 & ONS & ABG & CRO & LA & MWE \\ 
        2771 & 60404.9 & 25.5 & 7803 & 306.0 & 35.8 & 13625152.0 & ONS & UNG & CRO & OK & AOC \\ 
        2772 & 60424.3 & 19.4 & 300 & \textbf{\rd{15.5}} & 36.3 & 13439608.0 & ONS & ABG & CRO & IN & MWE \\ 
        2773 & 60458.0 & 33.8 & 227 & \textbf{\rd{6.7}} & 36.6 & 13319821.0 & ONS & ABG & CRO & MO & MWE \\ 
        2774 & 60475.6 & 17.6 & 5000 & 284.4 & 35.8 & 13607860.0 & ONS & ABG & CRO & TX & AOC \\ 
        2775 & 60494.2 & 18.6 & 2600 & 139.9 & 37.3 & 13070748.0 & ONS & ABG & CRO & TX & MWE \\ 
        2776 & 60539.9 & 45.8 & 906900 & 19823.0 & 36.2 & 13479949.0 & ONS & ABG & CRO & TX & COR \\ 
        2777 & 60565.3 & 25.3 & 26315 & 1038.8 & 36.6 & 13318231.0 & ONS & ABG & CRO & MO & MWE \\ 
        2778 & 60596.9 & 31.7 & 17456 & 551.2 & 35.8 & 13634449.0 & ONS & UNG & HFT & LA & AOC \\ 
        2779 & 60611.3 & 14.3 & 170 & \rd{\textbf{11.9}} & 37.1 & 13134282.0 & ONS & TAN & CRO & OK & MWE \\ 
        2780 & 60734.6 & 123.3 & 20333 & 164.9 & 37.0 & 13166475.0 & ONS & UNG & RPL & PA & AOC \\ 
        2781 & 60876.3 & 141.7 & 165593 & 1168.9 & 36.9 & 13205625.0 & ONS & ABG & CRO & TX & INO \\ \hline\hline
    \end{tabular}\label{tab:100fails}
\end{table}


\section{Conclusion} \label{sec:Con}
  
 This article introduced a risk-adjusted control chart to simultaneously monitor the time between and the cost associated with consecutive failures of systems. This is done by monitoring the ratio of the total cost (TC) and the time between events (TBE) variables based on an NHPP model.  The inclusion of risk adjustment serves to account for risk factors that improve the chart's sensitivity to variations in process failure and cost. The TBE and TC variables are assumed to be dependent such that a copula model can describe their dependency. The effectiveness of the proposed methods is assessed using the ARL (average run length) metric.  Extensive numerical simulations have been conducted to evaluate the performance of the monitoring procedures concerning various process parameters. In general, there are three categories of parameters, the parameters corresponding to the TBE ($\gamma$ and $\eta$) and TC ($\mu$) variables, the vector of coefficients $\pmb{\beta}$ associated with the risk model, and the parameter of the copula model $\theta_c$. During the simulation study, we only focused on assessing the performance of the proposed technique in monitoring $\gamma$, $\eta$ and $\mu$. In summary, the numerical study yields the following insights:
\begin{itemize}
\item[1.] The proposed chart is ARL-unbiased across all shift combinations, i.e., we always have $ARL_0 > ARL_1$.
\item[2.] The risk factors exhibiting a negative association with TBE have a positive impact on the performance of the proposed chart, while those with a positive association negatively impact the efficiency of the monitoring technique.
\item[3.]  Risk factors can notably increase the false alarm rate if not accounted for properly in the design of the chart.
\item[4.] A higher degree of dependency between TBE and TC improves the chart's ability to detect shifts sooner.
\end{itemize}

\section*{SupplementaryMaterials}

The online supplementary materials contain the R code and the dataset for the case study (the file \texttt{RCodesAndData.zip}).  The results of the case study can be reproduced using code and data in the following files:
\begin{itemize}
\item \texttt{data.csv}  
\item \texttt{R Script.R} 
\item \texttt{R Source.R} 
\end{itemize} 
Running the script file will reproduce the results of the case study. 

\section*{Data Availability Statement}

The dataset analyzed in the article is available as Supplementary materials.

\section*{Disclosure Statement}
The authors report there are no competing interests to declare.

\section*{Appendix: Estimation method}\label{Apn:MLE}
This section offers the required methodology to estimate the parameters of the model. Consider the continuous TBE and TC random variables, denoted as $X_i$ and $Y_i$ so that their joint distribution function is described by the copula function $C$. The joint PDF of TBE and TC is given in \eqref{equ:p.d.f}. Then, the likelihood function to estimate the unknown vectors $\bm{\theta}_X, \bm{\theta}_Y$, and $\theta_c$ can be calculated based on the paired observations $\left(x_i, y_i\right)_{i=1, \ldots, n}$ as:
\begin{align}
\notag &\ell\left(\bm{\theta}_X,\bm{\theta}_Y,\theta_c|\left(x_i, y_i\right)_{i=1, \ldots, n}  \right) \\
&=\sum_{i=1}^n \ln c\left(F_{\left(X_i|t_{i-1}, \bm{z_{i}}\right)}\left(x_i|\bm{\theta}_X\right),  F_Y\left(y_i|\bm{\theta}_Y\right)|\theta_c\right)+\sum_{i=1}^n \ln f_{\left(X_i|t_{i-1}, \bm{z_{i}}\right)}\left(x_i | \bm{\theta}_X\right)+\sum_{i=1}^n \ln f_Y\left(y_i |\bm{\theta}_Y\right),
\end{align}
where $c$ is copula pdf.  When dealing with multivariate models, it's common to encounter situations where closed-form estimators, such as maximum likelihood or other analytical methods, are not readily available. In such cases, numerical techniques become essential. Moreover, as the dimensionality of the problem increases, the challenges associated with numerical optimization become more pronounced, especially in the presence of a copula function. In light of these complexities, Joe and Xu (1996) proposed an approach called inference function for margins (IFM) which is particularly useful when we have data from multiple variables and we want to model their joint distribution using copulas. Joe and Xu (1996) showed that the IFM approach has some benefits over the traditional maximum likelihood estimation (MLE) method; it tends to be computationally simpler compared to MLE, especially for complex copula models and it can be more robust to misspecification of the marginal distributions. Even if the marginal models are not perfectly specified, the IFM approach can still provide reliable estimates of the copula parameters. In the current setting, one may encounter a problem with tens of risk factors, potentially significantly increasing the dimension of the parameter space. Therefore, we adopt the IFM approach to estimate the parameters. According to this approach, we apply the following two steps in the estimation procedure.
\begin{itemize}
\item[1.] Estimate the parameters of the margins $\bm{\theta}_X$ and $\bm{\theta}_Y$ as:
\begin{align} 
\hat{\bm{\theta}}_Y=\arg \max  \ell_Y\left(\hat{\bm{\theta}}_Y|\left(y_i\right)_{i=1, \ldots, n} \right),\\
\hat{\bm{\theta}}_X=\arg \max  \ell_X\left(\bm{\theta}_X|\left(x_i\right)_{i=1, \ldots, n} \right), \label{equ:thetax}
\end{align}
where $\ell_Y$ and $ \ell_X$ are the log-likelihood functions based $Y_i$ and $X_i|(t_{i-1},z_i)$ for $i=1,2,\ldots,n$, respectively. If $Y_i$ follows an Exponential distribution, like the case study, we have $\hat{\bm{\theta}}_Y=\mu$  and:
\begin{align}
\hat{\mu}=\frac{n}{\sum_{i=1}^ny_i}
\end{align}
\item[2.] Given $\hat{\bm{\theta}}_X$ and $\hat{\bm{\theta}}_Y$ obtained in step 1, estimate the copula parameter $\theta_c$ as:
\begin{align}
\hat{\theta}_c=\arg \max \ell_c\left( \theta_c,\hat{\bm{\theta}}_X, \hat{\bm{\theta}}_Y|\left(x_i, y_i\right)_{i=1, \ldots, n} \right),
\end{align}
where $\ell_c$ is the copula log-likelihood function.
\end{itemize}

To estimate the parameters using \eqref{equ:thetax}, the equation \eqref{equ:pdf_TBE} will be used to determined the log-likelihood function as:
\begin{align} \label{equ:logli}
\notag \ell_X(\bm{\theta}_X|x_1,x_2,...,x_n) &= \sum_{i=1}^n log \left(  \lambda\left(t_{i-1}+x_i| \bm{z_{i}}\right) \right) -\sum_{i=1}^n \left[ \Lambda\left(t_{i-1}+x_i| \bm{z_{i}}\right)-\Lambda\left(t_{i-1}| \bm{z_{i}}\right)\right] \\
\notag &=\sum_{i=1}^n log\left(\lambda(t_{i-1}+x_i) \exp \left(\bm{\beta}^{\prime} \bm{z}_i\right)\right) -\sum_{i=1}^n \big[\Lambda(t_{i-1}+x_i)  -\Lambda(t_{i-1}) \big]  \exp \big(\bm{\beta}^{\prime} \bm{z}_{i} \big)\\
&=\sum_{i=1}^n log(\lambda(t_{i-1}+x_i))+\sum_{i=1}^n \left(\bm{\beta}^{\prime} \bm{z}_i\right)  -\sum_{i=1}^n \big[\Lambda(t_{i-1}+x_i)  -\Lambda(t_{i-1}) \big]  \exp \big(\bm{\beta}^{\prime} \bm{z}_{i} \big).
 \end{align}

Accordingly, if we consider the power law intensity, then \eqref{equ:logli} reduces to:
 \begin{align} \label{equ:MLE-power-law}
\notag \ell_X\left( \bm{\beta}^{\prime}, \gamma, \eta |x_1, x_2,...,x_n \right)&=\sum_{i=1}^n log(\gamma \eta (t_{i-1}+x_i)^{\eta-1})+\sum_{i=1}^n \left(\bm{\beta}^{\prime} \bm{z}_i\right) \\
\notag & -\sum_{i=1}^n  \gamma \big((t_{i-1}+x_i)^{\eta} -t_{i-1}^{\eta}\big)   \exp \big(\bm{\beta}^{\prime} \bm{z}_{i}\big)\\&
\notag =n log(\gamma)+nlog(\eta)+(\eta-1)\sum_{i=1}^n log( t_{i-1}+x_i )
 \\&+\sum_{i=1}^n \left(\bm{\beta}^{\prime} \bm{z}_i\right) -\sum_{i=1}^n  \gamma \left[(t_{i-1}+x_i)^{\eta} -t_{i-1}^{\eta}\right]   \exp \big(\bm{\beta}^{\prime} \bm{z}_{i}\big)
 \end{align}

In the same way, in the case of the Log-Linear intensity, \eqref{equ:logli} reduces to:
  \begin{align} \label{equ:MLE-Log-Linear}
    \notag  \ell_X\left( \bm{\beta}^{\prime}, \gamma, \eta | x_1, x_2,...,x_n \right)&  = \sum_{i=1}^n log(exp(\gamma+\eta (t_{i-1}+x_i)))+\sum_{i=1}^n \left(\bm{\beta}^{\prime} \bm{z}_i\right) \\&
    \notag -\sum_{i=1}^n \frac{1}{\eta}\big(exp(\gamma+\eta (t_{i-1}+x_i))   -exp(\gamma+\eta t_{i-1}  ) \big)  \exp \left(\bm{\beta}^{\prime} \bm{z}_{i}\right)\\  
 \notag &=n\gamma+\eta \sum_{i=1}^n (t_{i-1}+x_i)+ \sum_{i=1}^n \left(\bm{\beta}^{\prime} \bm{z}_i\right) \\&
 -\frac{exp(\gamma)}{\eta} \sum_{i=1}^n \big[\exp( \eta (t_{i-1}+x_i))-exp( \eta t_{i-1}) \big] \exp(\bm{\beta}^{\prime} \bm{z}_{i}).
 \end{align}

Eventually, $\hat{\bm{\theta}}_X$ can be obtained in both scenarios by substituting \eqref{equ:MLE-power-law} and \eqref{equ:MLE-Log-Linear} in \eqref{equ:thetax} and solving the optimization problem.

\end{document}